\documentclass[12pt,a4paper]{article}
\usepackage{mathrsfs}
\usepackage{epsfig}
\usepackage{slashed}
\usepackage{graphicx}
\usepackage{subfig}
\usepackage{amssymb}
\usepackage{amsmath}
\usepackage{amsthm}
\begin{document}

\title{NLO QCD corrections to Single Top and W associated photoproduction at the LHC with forward detector acceptances}
\author{Hao Sun$^{1}$\footnote{haosun@mail.ustc.edu.cn \hspace{0.2cm} haosun@dlut.edu.cn},
        Wei Liu$^{1}$\footnote{liusd12@mail.dlut.edu.cn},
        Xiao-Juan Wang$^{1}$\footnote{wangxiaojuan@mail.dlut.edu.cn},
        Ya-Jin Zhou$^2$\footnote{zhouyj@sdu.edu.cn},
        Hong-Sheng Hou$^3$\footnote{hshou@hznu.edu.cn} \\
{\small $^{1}$ Institute of Theoretical Physics, School of Physics $\&$ Optoelectronic Technology,} \\
{\small Dalian University of Technology, Dalian 116024, P.R.China}\\
{\small $^{2}$ School of Physics, Shandong University, Jinan, Shandong 250100, P.R.China}\\
{\small $^{3}$ Department of Physics, Hangzhou Normal University, Hangzhou 310036, P.R.China}
}
\date{}
\maketitle

\vspace{-0.9cm}
\begin{abstract}

In this paper we study the Single Top and W boson associated photoproduction
via the main reaction $\rm pp\rightarrow p\gamma p\rightarrow pW^{\pm}t+Y$
at the 14 TeV Large Hadron Collider (LHC) up to next-to-leading order (NLO)
QCD level assuming a typical LHC multipurpose forward detector.
We use the Five-Flavor-Number Schemes (5FNS) with massless bottom quark
assumption in the whole calculation.
Our results show that the QCD NLO corrections can reduce the scale uncertainty.
The typical K-factors are in the range of 1.15 to 1.2 which lead to
the QCD NLO corrections of 15$\%$ to 20$\%$ correspond to the leading-order (LO)
predictions with our chosen parameters.

\vspace{-1.2cm} \vspace{2.0cm} \noindent
\end{abstract}

\newpage
\section{Introduction}

The Large Hadron Collider (LHC) at CERN generates high energetic proton-proton
($\rm pp$) collisions with a luminosity of $\rm {\cal L}=10^{34}cm^{-2}s^{-1}$
and provides the opportunity to study very high energy physics.
After the discovery of Higgs boson\cite{SMHiggs125GeV_ATLAS,SMHiggs125GeV_CMS},
probing new physics beyond the Standard Model (BSM) turns to the main goal of
the LHC. In such context, studying the heaviest elementary particle, the top quark,
is particularly interesting since it is the only fermion with a natural
Yukawa coupling to the Higgs boson of the order of unity.
Its charged weak coupling might be sensitive to the existence of an
additional heavy fermion. These couplings can be probed by measuring specific
top quark production cross sections and branching ratios.
However, these measurements will be challenging due to the composite internal
structure of the colliding particles, i.e., the large QCD or electroweak (EW)
backgrounds, the unknown precise centre-of-mass (c.m.s.) energy of the collisions
occurring between the partons of protons, the complicate composition of
underlying events within the central detector, etc.
In this case, very high energy interactions involving quasi-real incoming photons
may provide a solution to some of these problems.

General diagrams for the photon induced interactions at the LHC is presented
in Fig.\ref{rpexclusive}. $\rm pp\rightarrow p\gamma\gamma p\rightarrow pXp$
[left figure] refers to the photon-photon ($\gamma\gamma$) interaction where photons
radiated off by both protons collide and produce a central system X.
The system X will be detected by the central detector under clean experimental
conditions and the two protons remain intact (namely forward protons),
escape from the central detection and continue their path close to the beam line.
$\rm pp\rightarrow p\gamma p\rightarrow pXY$ [right figure] corresponds to photoproduction
or photon-proton ($\rm \gamma p$) production:
a photon from a proton induces a deep inelastic scattering
with the incoming proton and produces a proton remnant Y in addition to the
centrally produced X system. Despite a lower available luminosity, photoproduction
can occur under better known initial conditions, with fewer final states particles
and at high energy scale ($\rm \sim TeV$), thus can be studied as a complementary
tool to normal pp collisions at the LHC.
Indeed, the CDF collaboration has already observed such kinds of phenomenon
including the exclusive dilepton\cite{ppllpp1,ppllpp2}, diphoton\cite{pprrpp},
dijet \cite{ppjjpp} production and charmonium ($\rm J/\psi$) meson photoproduction\cite{ppJPHIpp}, etc.
Both the ATLAS and the CMS collaborations have programs of forward physics.
They are devoted to studies of high rapidity regions with extra updated detectors
located in a place nearly 100-400m close to the interaction point\cite{FDs1,FDs2,RevCEP}.
Technical details of the ATLAS Forward Physics (AFP) projects can be found,
for example, in Refs.\cite{AFP,AFP1}. A brief review of experimental prospects
for studying photon induced interactions are summarized in Ref.\cite{HEPhotonIntatLHC}.

As previously mentioned, the top quark is the heaviest known elementary particle
which makes it an excellent candidate for new physics searches.
Among top quark production channels, single top production has
some special features that top pair production can not achieve:
it offers a unique possibility of the direct measurement of
$\rm V_{tb}$, the Cabibbo-Kobayashi-Maskawa quark-mixing matrix (CKM),
allowing non-trivial tests of the properties of
this matrix in the SM\cite{SMCKMVtb1,SMCKMVtb2,SMCKMVtb3}.
In normal pp collision, single top produces mainly through two body
s-channel (t-channel) single top in association with a b (light) quark,
Wt channels and three body $\rm tbq'$ channel.
Here we focus on the study of Wt channel. This channel is invisible at the Tevatron,
however, it will be important at the LHC
and even comparable to single top s-channel production.
Even though, its cross section is almost a factor 100 smaller than the
most dangerous background coming from the $\rm t\bar{t}$ process.
This makes the measurement error of Wt process as large as $\sim 41\%$
for an integrated luminosity of $\rm 10\ fb^{-1}$ through normal pp collision\cite{wterrorLHC}.
Single top production can also proceed
through $\rm \gamma p$ collisions mentioned above.
This time through mainly Wt and $\rm tbq'$ channels.
Compare these two channels, we find that cross section of Wt channel
(order of $\rm \sim 1\ pb$), is much larger than that of $\rm tbq'$ channel (order of $\rm \sim 0.028\ pb$),
and becomes the most important single top production channel at the $\rm \gamma p$ collision at the LHC.
This is different from normal $\rm pp$ collision that
$\rm tbq'$ production channel accounts for $44.2\%$ ($39\%$)
of the total single top quark production cross section,
while Wt channel stands only $28\%$ ($5\%$) at LHC (Tevatron)\cite{WtLO_wbcut1}.
In contrast, Wt channel stands over $40\%$ of the top quark photoproductions,
since the top pair photoproduction has a cross section of only $\rm \sim1.5\ pb$.
Enhancement of the ratio $\rm \sigma_{Wt}/\sigma_{t\bar{t}}$ might certainly
be a good feature of related measurements through Wt channel.

First results on the measurement of the $\rm V_{tb}$ matrix element using Wt
photoproduction are presented in Refs.\cite{rbWtVtb1,rbWtVtb2}.
There comes the conclusion that the expected error on the measurement of
$\rm V_{tb}$ is $16.9\%$ for the semi-leptonic channel
and $10.1\%$ for the leptonic one after $\rm 10\ fb^{-1}$ of integrated luminosity,
while the expected uncertainty from the equivalent study
based on partonic interactions is $14\%$ \cite{gbWtVtb}
using the same integrated luminosity, showing that
photoproduction is at least competitive with partonic-based studies
and that the combination of both studies could lead to significant
improvement of the error.

In addition to the $\rm V_{tb}$ measurement, Wt photoproduction
can also be used to study the W-t-b vertex and test precisely the
V-A structure of the charged current weak interaction of the top quark.
Anomalous measurement of this vertex may lead direct evidence of new physics
beyond the SM. It may manifest itself via either loop effects or inducing
non-SM weak interactions to introduce new single top production channels.
Typical studies involve, i.e., measuring anomalous W-t-b coupling
in ep collision\cite{epWtbAno1,epWtbAno2},
in normal pp collision\cite{rbWtbVA1,rbWtbVA2,gbWtnewPhysics}
as well as in $\rm \gamma p$ collision at the LHC\cite{AnomalousWtb}.
Ref.\cite{AnomalousWtb} studies $\rm pp\rightarrow p\gamma p \rightarrow pW^{\pm}t+Y$
up to the leading order (LO) induced by anomalous W-t-b coupling.
In this case, SM Wt photoproduction turns to its irreducible background.
Moreover, a lot of studies are performed at the $\rm \gamma p$ colliders,
i.e., testing anomalous gauge boson couplings
\cite{rpanoWWr1,rpanoWWr2,rpanoVVV,rpanoZZrr,rpanoZZrZrr,rpanoWWrrZZrrrp}
or probing flavour changing neutral currents (FCNC)
through single top photoproduction\cite{Anomaloustqr,Anomaloustqr1}, etc.
In cases like these, SM Wt photoproduction turns to be a most important reducible background
that needs precise measurement and analysis.
Furthermore, determining Wt production cross sections and once
compared with experiments, will provide a direct access
to the bottom quark parton density in nucleons and help understand
the nature of the b quark parton distribution function (PDF).

As a result, accurate theoretical predictions including higher order
QCD corrections for Wt photoproduction are needed.
NLO QCD corrections for Wt production in the normal pp collisions
have already been well studied
\cite{WtNLO1,WtNLO2,WtNLO3,WtNLO4,WtNLO5,WtNLO_NewMethod,WtNLO_bPDF,WtNLO6}.
In this paper, we present its production at the $\rm \gamma p$ collision
for the first time, assuming a typical LHC multipurpose forward detector,
including the NLO QCD corrections
via the main reaction $\rm pp\rightarrow p\gamma p\rightarrow pW^{\pm}t+Y$.
Typically, we use the Five-Flavor-Number Schemes (5FNS) with massless b quark mass assumption
through the whole calculation. Our paper is organized as follow:
we build the calculation framework in Section 2 including brief
introduction to the Equivalent Photon Approximation (EPA),
general inelastic photoproduction cross section, LO and QCD NLO
Wt photoproductions. Section 3 is arranged to present the
input parameters, cross checks and numerical results of our study.
Finally we summarize our conclusions in the last section.

\section{Calculation Framework}

\subsection{Equivalent Photon Approximation}

\begin{figure}[hbtp]
\vspace{-4.0cm}
\hspace*{-2cm}
\includegraphics[scale=0.8]{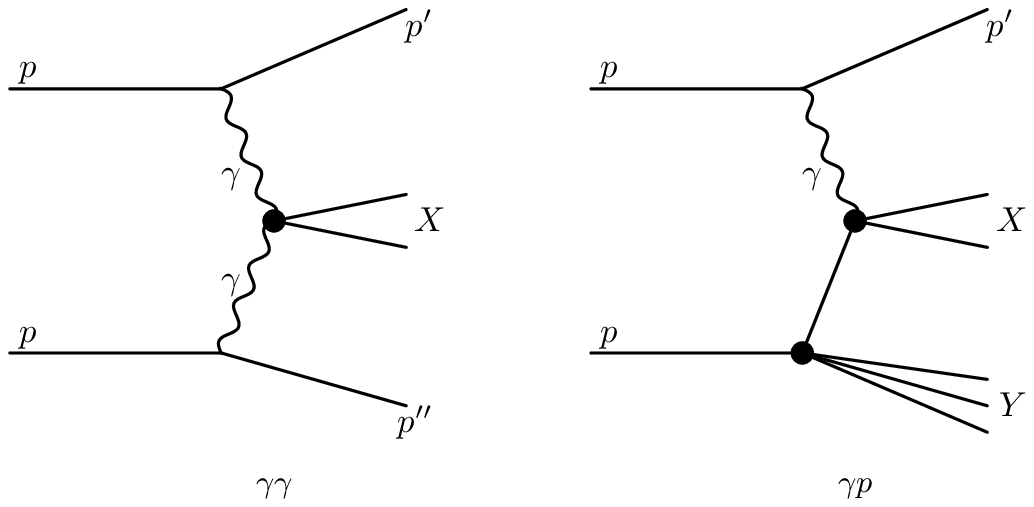}
\vspace{-16cm}
\caption{\label{rpexclusive}
Generic diagrams for the photon induced production at the CERN LHC:
$\rm pp\rightarrow p\gamma\gamma p\rightarrow pXp$ [left figure] and
$\rm pp\rightarrow p\gamma p\rightarrow pXY$ [right figure].}
\end{figure}

In our paper, we focus on the discussion of photoproductions
$\rm pp\rightarrow p\gamma p\rightarrow pXY$ through $\rm \gamma p$ collisioins
{see Fig.\ref{rpexclusive}[right figure]}.
Photoproduction is a class of processes in which one of the two interacting protons is
not destroyed during the collision but survive into the final state with additional particle
(or particles) state(s). Protons of this kind are named intact or forward protons.
The kinematics of a forward proton is often described by means of the reduced energy loss $\xi$,
which is also defined as the forward detector acceptance:
\begin{eqnarray}\label{Eq.xi}
 \xi=\rm \frac{\Delta E}{E}=\frac{E-E'}{E}
\end{eqnarray}
where E is the initial energy of the beam and $\rm s=4 E^2$ is the square of the centre-of-mass energy.
$\rm E'$ is the energy after the interaction and $\Delta E$ is the energy that the proton lost in the interaction.
Compare to the usual pp Deep Inelastic Scattering (DIS), $\gamma\gamma$ and $\rm \gamma p$
collisions can provide more clean environment. Compare themselves, $\gamma\gamma$ collisions
can be cleaner than the $\rm \gamma p$ collisions. However, $\rm \gamma p$ collisions
have higher energy and effective luminosity with respect to $\gamma\gamma$ collisions.

Processes through $\gamma\gamma$ or $\rm \gamma p$ interactions
involve photon exchange with proton beams at the LHC
which can be described by the appropriate framework of
equivalent photon (or Weizs$\rm \ddot{a}$cker-Williams) approximation (EPA) \cite{EPA}.
In the framework of EPA, emitted quasi-real photons
from the protons have a low virtuality and scattered
with small angles from the beam pipe.
Therefore the emitters proton should also be scattered
with a small angle and tagged by the forward detectors
with some momentum fraction loss $\xi$ given in Eq.(\ref{Eq.xi}).
Higher $\xi$ can be obtained with the closer installation
of the forward detectors from the interaction points.
The emitted quasi-real photons by the emitters protons
with small angles show a spectrum of virtuality $\rm Q^2$
and the energy $\rm E_\gamma$. This is described by the EPA
which differs from the point-like electron (positron) case
by taking care of the electromagnetic form factors in the equivalent
photon spectrum and effective photon luminosity:
\begin{equation}\label{EPAformula}
\rm \frac{dN_\gamma}{dE_\gamma dQ^2}=\rm
\frac{\alpha}{\pi}\frac{1}{E_\gamma Q^2}\left[(1-\frac{E_\gamma}{E})(1-\frac{Q^2_{min}}{Q^2})F_E
 + \frac{E^2_\gamma}{2 E^2}F_M \right]
\end{equation}
with
\begin{eqnarray} \nonumber
\rm Q^2_{min}=( \frac{M^2_{inv} E}{E-E_{\gamma}} -M^2_P )\frac{E_{\gamma}}{E},
 ~~~~ F_E= \frac{4 M^2_p G^2_E + Q^2 G^2_M}{4 M^2_p +Q^2}, \\
\rm G^2_E=\frac{G^2_M}{\mu^2_p}=(1+\frac{Q^2}{Q^2_0})^{-4}, ~~~~F_M=G^2_M, ~~~~Q^2_0=0.71 GeV^2 ,
\end{eqnarray}
where $\rm \alpha$ is the fine-structure constant, E is the energy of the incoming proton beam.
which is related to the quasi-real photon energy by $\rm E_\gamma=\xi E$.
$\rm M_p$ is the mass of the proton and $\rm M_{inv}$ is the invariant mass of the final state.
$\rm \mu^2_p$ = 7.78 is the magnetic moment of the proton.
$\rm F_E$ and $\rm F_M$ are functions of the electric and magnetic form factors
given in the dipole approximation.

Many phenomenological studies on photon induced processes are summarized here involve:
standard model productions \cite{SMWH},
supersymmetry\cite{SUSYprrp1,SUSYprrp2},
extra dimensions\cite{EDpllp1,EDprrp2,EDprrp3,EDrqrq4},
unparticle physics\cite{unparticle},
top triangle moose model\cite{TTMrbtp},
gauge boson self-interactions
\cite{rpanoWWr1,rpanoWWr2,rpanoVVV,rpanoZZrr,rpanoZZrZrr,rpanoWWrrZZrrrp,rranoWWr3,rranoWWrr,rranoZZZ,rranoVVVV},
neutrino electromagnetic properties\cite{electromagnetic1,electromagnetic2,electromagnetic3},
the top quark physics\cite{rbWtVtb1,rbWtVtb2,AnomalousWtb,Anomaloustqr,Anomaloustqr1},
dark matter searches\cite{rpDMj}
and triplet Higgs production\cite{TripletH}, etc.

\subsection{General $\rm \gamma p$ Photoproduction Cross Section}

We denote the general photoproduction processes at the LHC,
no metter at LO or NLO level,  as
\begin{eqnarray}
\rm pp \rightarrow  p\gamma p \rightarrow p + \gamma + q/\bar{q}/g
\rightarrow p + \underbrace{i + j + k + ...}_X +  Y
\end{eqnarray}
with q = u, d, c, s, b and i, j, k, ... the final state particles.
The hadronic cross section at the LHC can be converted by integrating
$\rm \gamma + q/\bar{q}/g \rightarrow i + j + k +...$ over the photon ($\rm dN(x,Q^2)$),
gluon and quark ($\rm G_{g,q/p}(x_2,\mu_f)$) spectra:
\begin{eqnarray}\label{totcrosssection}  \nonumber
\rm \sigma_{\gamma p}=\rm \sum_{j=q,\bar{q},g} \int^{\sqrt{\xi_{max}}}_{\frac{M_{inv}}{\sqrt{s}}} 2z dz \int^{\xi_{max}}_{Max(z^2,\xi_{min})}
\frac{dx_1}{x_1} \int^{Q^2_{max}}_{Q^2_{min}} \frac{dN_\gamma(x_1)}{dE_\gamma dQ^2} G_{g,q/p}(\frac{z^2}{x_1}, \mu_f) \\
\rm \cdot \int \frac{1}{avgfac} \frac{|{\cal M}_n (\gamma j\rightarrow klm..., \hat s =z^2 s )|^2}{2 \hat s (2 \pi)^{3n-4}} d\Phi_n ,
\end{eqnarray}
where $\rm x_1$ is the ratio between scattered quasi-real photons and incoming proton energy
$\rm x_1 = E_\gamma/E$. $\rm \xi_{min} (\xi_{max})$ are its lower (upper) limits which means that
the forward detector acceptance satisfies $\rm \xi_{min}\leq\xi\leq\xi_{max}$.
$\rm x_2$ is the momentum fraction of the proton momentum carried by the gluon (quark).
The quantity $\rm \hat s = z^2 s$ is the effective c.m.s. energy with
$\rm z^2=x_1 x_2$. $\rm s=4 E^2$ mentioned above and $\rm M_{inv}$ is the total mass
of the related final states. $\rm 2z/x_1$ is the Jacobian determinant
when transform the differentials from $\rm dx_1dx_2$ into $\rm dx_1dz$. $\rm G_{g,q/p}(x,\mu_f)$
represent the gluon (quark) parton density functions, $\rm \mu_f$ is the factorization scale.
$\rm f=\frac{dN}{dE_\gamma dQ^2}$
is the $\rm Q^2$ dependent relative luminosity spectrum present in Eq.(\ref{EPAformula}).
$\rm Q^2_{max}=2 GeV^2$ is the maximum virtuality.
$\rm \frac{1}{avgfac}$ is the product of the spin-average factor, the color-average factor
and the identical particle factor. $\rm |{\cal M}_n|^2$ presents the squared n-particle matrix element
and divided by the flux factor $\rm [2 \hat s (2 \pi)^{3n-4}]$. The n-body phase space differential
$\rm d\Phi_n$ and its integral $\rm \Phi_n$ depend only on $\rm \hat s$
and particle masses $\rm m_i$ due to Lorentz invariance:
\begin{eqnarray} \nonumber
\rm \Phi_n(\hat s, m_1, m_2, ..., m_n) &=&\rm  \int d\Phi_n(\hat s, m_1, m_2,..., m_n) \\
&=&\rm  \int \delta^4((p_i+p_j)-\sum^{n}_{k=1}p_k) \prod^{n}_{k=1}d^4 p_k \delta(p^2_k-m^2_k) \Theta (p^0_{k})
\end{eqnarray}
with i and j denoting the incident particles and k running over all outgoing particles.

\subsection{Wt Photoproduction at Leading Order}

\begin{figure}[hbtp]
\vspace{-5cm}
\hspace*{-3cm}
\includegraphics[scale=0.9]{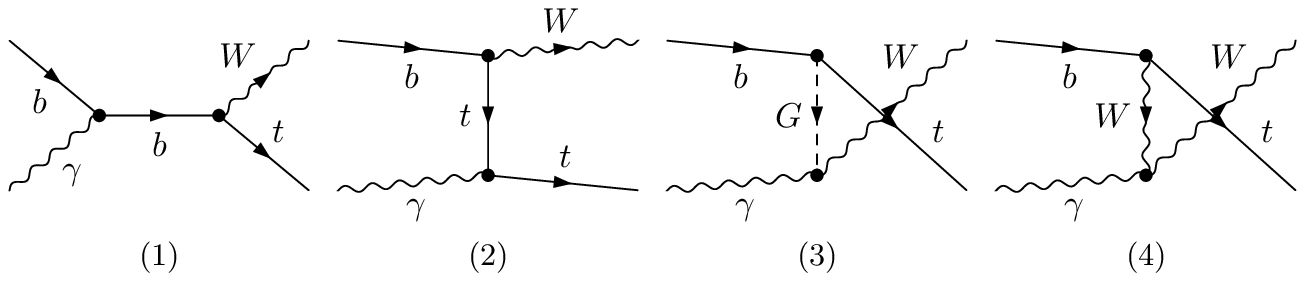}
\vspace{-20cm}
\caption{\label{figrbwtborn} Tree level Feynman diagrams for $\rm \gamma b \rightarrow W^-t$ in the SM.}
\end{figure}

We denote the Wt photoproduction process as:
\begin{eqnarray}
\rm pp\rightarrow p\gamma p\rightarrow p\gamma(p_1)b(p_2) \rightarrow p W^-(p_3)t(p_4) +Y
\end{eqnarray}
where $\rm p_i$ are the particle four momentums.
There are four LO Feynman diagrams for
this partonic process as shown in Fig.\ref{figrbwtborn}.
There Fig.\ref{figrbwtborn}(1) and Figs.\ref{figrbwtborn}(2-4)
are the s-channel and t-channel diagrams for the partonic
process, respectively. Fig.\ref{figrbwtborn}(3) include
b-t-G vertex that can be safely omitted in the massless b
quark assumption. We only consider the
$\rm W^-t$ production while its charge-conjugate contribution
is the same \cite{WtNLO_NewMethod}.
In order to describe the process
$\rm \gamma(p_1) b(p_2) \rightarrow W^-(p_3) t(p_4)$ we choose the c.m.s.
($\rm \textbf{p}_1+\textbf{p}_2=0$) with the momentum directed along z-axis.
In c.m.s. the particle momentum read
\begin{eqnarray} \label{Eq.notationE}\nonumber
\rm p_1&=&\rm \frac{\sqrt{\hat s}}{2}(e_1,0,0,e_{1z}),  \ \ \ p_3= \frac{\sqrt{\hat s}}{2}(e_3,e_{3x},e_{3y},e_{3z}) \\ \rm p_2&=&\rm \frac{\sqrt{\hat s}}{2}(e_2,0,0,e_{2z}),\ \ \ p_4= \frac{\sqrt{\hat s}}{2}(e_4,e_{4x},e_{4y},e_{4z}).
\end{eqnarray}
Here notation of $\rm e_{i/ix}$ equal $\rm p_{i/ix}/(\sqrt{\hat s}/2)$
and is needed in our following description.

The LO cross section for the partonic process $\rm \gamma b\rightarrow W^- t$
is obtained by using the following formula
\begin{eqnarray}
\rm \hat\sigma^{LO}(\hat{s},\gamma b\rightarrow W^- t)=
 \frac{(2\pi)^4}{4|\textbf{p}_1|\sqrt{\hat{s}}} \int \overline\sum|{\cal M}^{LO}|^2 d\Phi_2
\end{eqnarray}
where $\rm d\Phi_2$ is the two-body phase-space element,
and $\rm \textbf{p}_1$ is the momentum of the initial photon
in the c.m.s.. The integration
is performed over the two-body phase space of the final particles $\rm W^- t$.
The summation is taken over the spins and colors of the initial and
final states, and the bar over the summation indicates averaging over the
intrinsic degrees of freedom of initial partons.

The LO total cross section for $\rm pp\rightarrow p\gamma p\rightarrow pW^- t$
can be expressed as
\begin{eqnarray}\label{Eq.wtLO}\nonumber
&&\rm \sigma^{LO} (pp\rightarrow p\gamma p \rightarrow pW^-t+Y)\\\nonumber
&=&\rm  \int^{\sqrt{\xi_{max}}}_{\frac{M_{inv}}{\sqrt{s}}} 2z dz \int^{\xi_{max}}_{Max(z^2,\xi_{min})}
\frac{dx_1}{x_1} f_{\gamma/P_A}(x_1)   G_{b/P_B}(\frac{z^2}{x_1}, \mu_f) \hat\sigma^{LO} (\gamma b\rightarrow W^-t, z^2s,\mu_f,\mu_r)\\
&+&\rm (A\leftrightarrow B).
\end{eqnarray}
There $\rm G_{i/P_{j}}$, i=b, j = A, B represent the PDFs of parton i
in proton $\rm P_{j}$ , $\rm \mu_f$ and $\rm \mu_r$ are the factorization and
renormalization scales separately.
Here we use $\rm f_{\gamma/P_A} (x_1)$ to take place of the
$\rm \int^{Q^2_{max}}_{Q^2_{min}} \frac{dN_\gamma(x_1)}{dE_\gamma dQ^2}$ in Eq.(\ref{totcrosssection})
for simplicity. And we address here that during calculation,
we use the $\xi$, $\rm Q^2$ dependent form of Eq.(\ref{totcrosssection}).

\subsection{Wt Photoproduction at QCD Next-to-Leading Order}

\subsubsection{General Description}

We use the Five-Flavor-Number Scheme (5FNS) in our whole LO and
QCD NLO calculations. As we see, at tree level in the 5FNS scheme
the Wt photoproduction process consists of only one partonic subprocess,
namely $\rm \gamma b\rightarrow W^-t$, as illustrated in Fig.\ref{figrbwtborn}.
Indeed, Wt photoproduction can also be produced in the
Four-Flavor-Number Scheme (4FNS) where the b quark is treated as massive
and there is no b quark parton density is assumed in the initial state.
In this scheme, the LO contribution starts from $\rm \gamma g\rightarrow W^-t\bar{b}$
with 1b tagged in the final state. The first order of QCD corrections in 4FNS
consist of virtual one-loop corrections to the tree-level subprocesses as
well as real corrections in the form of other two subprocesses with an additional
radiated parton, namely $\rm \gamma g\rightarrow W^-t\bar{b}+g$ and
$\rm \gamma q\rightarrow W^-t\bar{b}+\bar{q}$.
In the 4FNS scheme, b quark do not enter in the computation of the running of $\rm \alpha_s$
and the evolution of the PDFs. Finite $\rm m_b$ effects enter via power corrections
of the type $\rm (m^2_b/Q^2)^n$ and logarithms of the type $\rm log^n (m^2_b/Q^2)$
where Q stands for the hard scale of the process. At the LHC, typically $\rm (m_b/Q) \ll 1$
and power corrections are suppressed, while logarithms, both of initial and final state
nature, could be large. These large logarithms could in principle spoil the convergence of fixed order
calculations and a resummation could be required. Up to NLO accuracy those potentially large
logarithms, $\rm log(m_b/Q)$, are replaced by $\rm log(p^{min}_{T,b}/Q)$ with $\rm m_b \ll p^{min}_{T,b} \leq Q$
and are less significant numerically. As can be see, the difference between adopting the 5FNS
and 4FNS is the ordering of the perturbative series for the production cross section.
In the 4FNS the perturbative series is ordered strictly by powers of the strong coupling
$\rm \alpha_s$, while in the 5FNS the introduction of the b quark PDF allows to resum
terms of the form $\rm \alpha^n_s log(\mu^2/m^2_b)^m$ at all orders in $\rm \alpha_s$.
If all orders in perturbation theory were taken into account,
these two schemes are identical in describing logarithmic effects.
But the way of ordering in the perturbative expansion is different and
at any finite order the results might be different.
Many works have been done in the comparison of the 5FNS and 4FNS Schemes,
see Refs.\cite{45FNSa,45FNSb,45FNSc,45FNSd,45FNSe,45FNSf,45FNSg}, etc.
A latest comparison in the 4FNS and 5FNS schemes in Ref.\cite{45FNS}
present that being often the effects of resummation very mild,
4FNS calculations can be put to use, on the other hand, for 5NFS schemes, i.e.,
can typically provide quite accurate predictions for total rates and being simpler,
in some cases allow the calculations to be performed at NNLO.
We address the interesting of considering both schemes while here we use 5FNS in our
calculation. Even in 5FNS, it will be interesting to consider two schemes
\cite{massiveNLO1,massiveNLO2}:
one is the massless b quark scheme where we drop the mass of b quark during calculation
while the other is the massive b quark scheme where we retain it everywhere.

In our paper, we adopt the 5FNS scheme with the massless b quark assumption.
In this case, the first order of QCD
corrections to the $\rm pp\rightarrow p\gamma p \rightarrow pW^-t+Y$ consist of:
\begin{itemize}
\item The QCD one-loop virtual corrections to the partonic process $\rm \gamma b\rightarrow W^-t$.
\item The contribution of the real gluon radiation partonic process $\rm \gamma b\rightarrow W^- t + g$.
\item The contribution of the real b-quark emission partonic process $\rm \gamma g\rightarrow W^- t + \bar{b}$.
\item The corresponding contributions of the PDF counterterms.
\end{itemize}

We use the dimensional regularization method in $\rm D=4-2\epsilon$ dimensions
to isolate the ultraviolet (UV) and infrared (IR) singularities.
In massless b quark scheme, we split each collinear counter-term of the PDF,
$\rm \delta G_{b/P}(x,\mu_f)$ (P=proton), into two parts:
the collinear gluon emission part $\rm \delta G^{gluon}_{b/P}(x,\mu_f)$
and the collinear b quark emission part $\rm \delta G^{quark}_{b/P}(x,\mu_f)$.
The analytical expressions are presented as follows
\begin{eqnarray}
\rm \delta G_{b/P}(x,\mu_f) = \delta G^{gluon}_{b/P}(x,\mu_f) + \delta G^{quark}_{b/P}(x,\mu_f)
\end{eqnarray}
with
\begin{eqnarray} \label{Eq.masslessb} \nonumber
\rm \delta G^{gluon}_{b/P} &=& \rm \frac{1}{\epsilon}
   \left[\frac{\alpha_s}{2\pi}\frac{\Gamma(1-\epsilon)}{\Gamma(1-2\epsilon)}
   \left(\frac{4\pi\mu^2_r}{\mu^2_f}\right)^{\epsilon}\right]
   \int^1_{x} \frac{dz}{z} P_{bb}(z) G_{b/P}(\frac{x}{z},\mu_f) \\
\rm \delta G^{quark}_{b/P} &=&  \rm \frac{1}{\epsilon}
   \left[\frac{\alpha_s}{2\pi}\frac{\Gamma(1-\epsilon)}{\Gamma(1-2\epsilon)}
   \left(\frac{4\pi\mu^2_r}{\mu^2_f}\right)^{\epsilon}\right]
   \int^1_{x} \frac{dz}{z} P_{bg}(z) G_{g/P}(\frac{x}{z},\mu_f)
\end{eqnarray}
and the explicit expressions for the splitting functions $\rm P_{ij}(z)$, (ij = bb, bg) can be found
in Ref.\cite{2PSS:Owens}.

\subsubsection{Virtual}

\begin{figure}[hbtp]
\vspace{-5cm}
\hspace*{-3cm}
\centering
\includegraphics[scale=0.8]{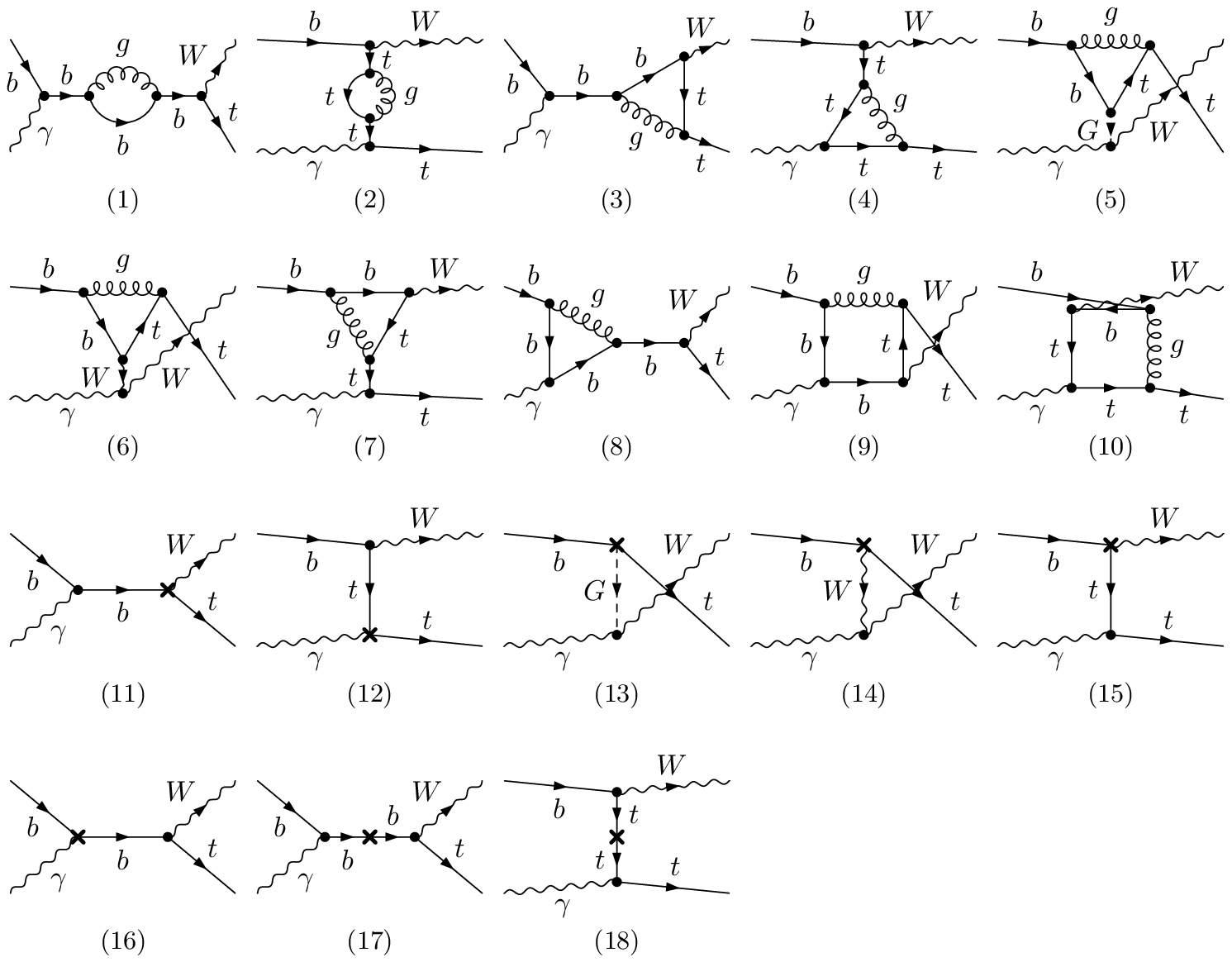}
\vspace{-9cm}
\caption{\label{figrbwtloop} QCD one loop Feynman diagrams for $\rm \gamma b \rightarrow W^-t$ in the SM.}
\end{figure}

The amplitude at the QCD one-loop level for the partonic process
$\rm \gamma b\rightarrow W^- t$ in the SM contains the contributions of
the self-energy, vertex, box and counter-term graphs which are shown
in Fig.\ref{figrbwtloop}(1)-(18). Same as the leading level
that diagrams include b-t-G vertex that can be safely omitted in the
massless b quark scheme. Even in the massive b quark scheme, their contributions
are also quite small.

To remove the UV divergences, we need to renormalize the mass of the quarks
and the wave function of the quark fields. In the massless b quark assumption
we introduce the following renormalization constants:
\begin{eqnarray}\nonumber
\rm \psi^{0,L,R}_{b(t)}&=&\rm \left(1+\delta Z^{L,R}_{\psi_{b(t)}}\right)^{\frac{1}{2}} \psi^{L,R}_{b(t)}, \\
\rm m^0_t &=&\rm m_t+\delta m_t,
\end{eqnarray}
where $\rm m_t$ are the top-quark mass. $\rm \psi^{L,R}_{b(t)}$ denote the fields of bottom (top) quark.
For the masses and wave functions of the fields
are renormalized in the the on-shell scheme and the relevant counter-terms
are expressed as
\begin{eqnarray} \nonumber
\rm \ \delta Z^{L,R}_{\psi_{t}} &=&\rm  -\frac{\alpha_{s}}{4\pi}C_F\left[\Delta_{UV}+ 2 \Delta_{IR} +4
                                              + 3ln\left(\frac{\mu^2_r}{m^2_t}\right) \right], \\
\rm \frac{\delta m_t}{m_t} &=&\rm  -\frac{\alpha_s}{4\pi} C_F \left[3 \Delta_{UV} + 4
                                              + 3ln\left(\frac{\mu^2_r}{m^2_t}\right) \right],
\end{eqnarray}
with $\rm C_{F}=\frac{4}{3}$, $\rm \Delta_{UV(IR)}=\frac{1}{\epsilon_{UV(IR)}}
\Gamma(1+\epsilon_{UV(IR)})(4\pi)^{\epsilon_{UV(IR)}}$
refer to the $\rm UV(IR)$ divergences.
For massless b quark, there is no need to renormalize the mass of bottom
and we use modified minimal subtraction ($\rm \overline{MS}$) scheme for b field as:
\begin{eqnarray}
\rm \delta Z^{L,R}_{\psi_{b}} &=&\rm -\frac{\alpha_{s}}{4\pi}C_F \left[\Delta_{UV}-\Delta_{IR}\right].
\end{eqnarray}
UV singularities are regulated by adding renormalization part to the virtual
corrections only leaving IR singularities that will be removed by combining
the real emission corrections.
We calculate the virtual one-loop corrections ($\rm \sigma^{V}$) using a Feynman diagram
approach based on FeynArts, FormCalc and our modified LoopTools
(FFL)\cite{FeynArts,FormCalc,LoopTools} packages. Tensor one loop integrals are
checked with OneLoop\cite{Oneloop} and QCDLoop\cite{QCDloop} packages.

\subsubsection{Parton Radiation}

\begin{figure}[hbtp]
\vspace{-4.5cm}
\hspace*{-1.8cm}
\includegraphics[scale=0.8]{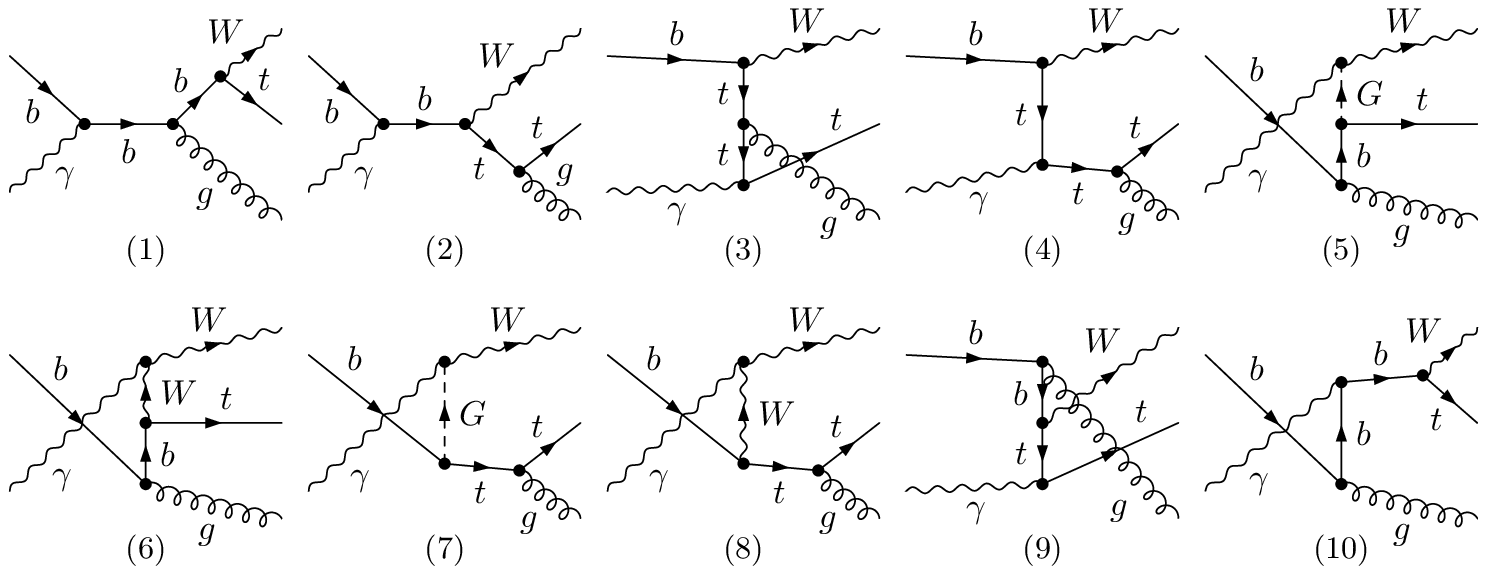}
\vspace{-14cm}
\caption{\label{rbwthardg}
The tree parton level Feynman diagrams for
the real gluon emission subprocess
$\rm \gamma b \rightarrow W^- t g$ related to Eq.(\ref{Eq.hard1}).}
\end{figure}

\begin{figure}[hbtp]
\vspace{-5cm}
\hspace*{-1.8cm}
\includegraphics[scale=0.8]{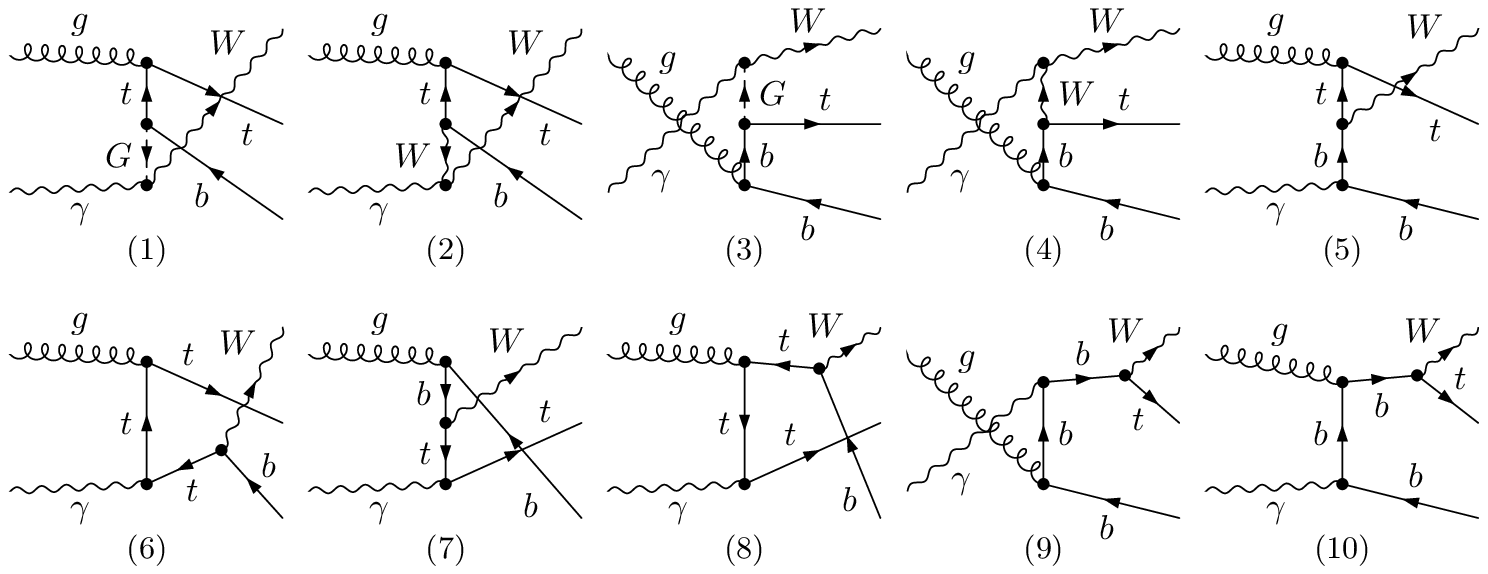}
\vspace{-14cm}
\caption{\label{rbwthardbbar}
The tree parton level Feynman diagrams for
the real light-(anti)quark emission subprocess
$\rm \gamma g\rightarrow W^- t \bar{b}$ related to Eq.(\ref{Eq.hard2}).}
\end{figure}

The first order of QCD corrections also consist of the real corrections
in the form of other two subprocesses with an additional
radiated parton, namely gluon emission and real (anti) quark emission presented as
\begin{eqnarray}\label{Eq.hard1}
&&\rm \ \ \gamma(p_1) b(p_2) \rightarrow W^-(p_3) t(p_4) g(p_5) \\\label{Eq.hard2}
&&\rm \ \ \gamma(p_1) g(p_2) \rightarrow W^-(p_3) t(p_4) \bar{b}(p_5)
\end{eqnarray}
with Feynman diagrams depicted in Fig.\ref{rbwthardg} and
Fig.\ref{rbwthardbbar}, respectively.

In the massless b quark scheme, singularities associated with
initial state collinear gluon emission are absorbed into the definition
of the PDFs, see in Eq.(\ref{Eq.masslessb}).
We employ the $\rm \overline{MS}$ scheme for the parton
distribution functions. Similar to the virtual part, we utilize
dimensional regularization (DR) to control the singularities of the
radiative corrections, which are organized using the two cutoff
phase space slicing (TCPSS) method\cite{2PSS:Owens}.
Since we treat the b quark as massless, there is collinear IR
singularity which is regularized by $1/\epsilon$ in the DR scheme.
This term is canceled by the corresponding contribution in the b quark
PDF counterterm, in other words, absorbed by the b quark PDF.
This cancellation has been checked both analytically and numerically,
therefore, avoid double counting problem.
We adopt TCPSS to isolate the IR singularities by introducing
two cutoff parameters $\rm \delta_{s}$ and $\rm \delta_{c}$. An arbitrary
small $\rm \delta_{s}$ separates the three-body final state phase space
into two regions: the soft region ($\rm E_{5}\leq \delta_{s}\sqrt{\hat{s}}/2$)
and the hard region ($\rm E_{5}>\delta_{s}\sqrt{\hat{s}}/2$).
The $\rm \delta_{c}$ separates hard region into the hard collinear
(HC) region and hard noncollinear ($\rm \overline{HC}$) region.
The criterion for separating the HC region is described as follows:
the region for real gluon/light quark emission with $\rm \hat{s}_{15}$
(or $\rm \hat{s}_{25}) < \delta_{c}\hat{s}$ (where
$\rm \hat{s}_{ij}=(p_{i}+p_{j})^{2}$) is called the HC region.
Otherwise it is called the $\rm \overline{HC}$ region.

At QCD NLO, some of the contributions representing the emission of
an additional parton require special attention. For example,
when we calculate the remain part in real radiation corrections
$\rm \gamma g\rightarrow Wtb$ (Eq.\ref{Eq.hard2}),
appropriate crossing of the diagrams shown in Fig.\ref{rbwthardbbar}
should be included. Some of the diagrams which produce a final state consisting
of a W, an on-shell top quark and a b quark are particularly problematic.
During phase space integration, one need to integrate over the region $\rm M^2_{Wb}=(p_W+p_b)$.
If in this case, a resonant t propagator (with flowing momentum equal or close to $\rm p_W+p_b$)
is encountered, a divergence will arise. Actually these diagrams can be interpreted
as the production of a $\rm t\bar{t}$ pair at LO, with subsequent decay of
the top into a Wb system. This is the well known interference between Wt and
$\rm t\bar{t}$ production, namely doubly resonant. Such resonant becomes extremely large
in certain phase space region and renders the perturbative computation of the
Wt cross section meaningless, thus should be preferable excluded from the
NLO corrections to the Wt process.

Several approaches have been outlined in the literature, i.e.,
making a cut on the invariant mass of the Wb system to prevent the t propagator
from becoming resonant \cite{WtLO_wbcut1}, subtracting the contribution from the
resonant diagrams so that no on-shell piece remains \cite{WtLO_wbcut2,WtLO_wbcut3},
bottom quark PDF method, technically, perform calculation of the Wt
process by applying a veto on the $\rm p_T$ of the additional b quark that appears
at next-to-leading order aids the separation of this process from doubly-resonant
$\rm t\bar{t}$ production \cite{WtNLO_bPDF}, etc \cite{WtNLO_NewMethod,WtLO_wbcut4}.
Here in our case we use the PROSPINO scheme \cite{Prospino} which is defined as
a replacement of the Breit-Wigner propagator
\begin{eqnarray} \nonumber
\rm \frac{| {\cal M}|^2(s_{Wb})}{(s_{Wb}-m^2_t)^2+ m_t^2 \Gamma_t^2} \rightarrow
&&\rm \frac{|{\cal M}|^2(s_{Wb})}{(s_{Wb}- m^2_t)^2+ m_t^2 \Gamma_t^2}- \\
&&\rm \frac{|{\cal M}|^2(m^2_t)}{(s_{Wb}-m^2_t)^2+ m_t^2 \Gamma_t^2} \Theta (\hat{s}-4  m_t^2) \Theta (m_t-m_W).
\end{eqnarray}
This subtraction scheme helps to avoid double counting and to not artificially
ruin the convergence of the perturbative QCD description of these production channels
with the remove of on-shell particle contributions from the associated production.
This scheme has be done in some other refs
like \cite{usePROSPINO1, usePROSPINO2, usePROSPINO3, usePROSPINO4}, etc.

Some attention should be paid to the light-(anti)quark emission subprocess, 
see Figs.5(5) and Fig.5(10). Contributions from these two diagrams can be considered
as part of the NLO EW corrections to the LO process $\rm pp\rightarrow gb\rightarrow Wt$
through normal pp collision. 
Since we concentrate on the photoproduction of $\rm pp\rightarrow p\gamma p\rightarrow pW^-t$ 
where forward protons are considered, 
the $\rm pp\rightarrow gb\rightarrow Wt$ and its full NLO EW corrections are not taken into account.
Then, the subprocesses in Figs. 5(5) and 5(10) 
are defined applying a small $\rm p_T$ cut on the tagged b quark, 
which regularize the collinear splitting of the photon into a $\rm b\bar{b}$ pair.
Indeed, the choice of specific kinematical cuts to select events in the forward region
(small $\rm p_T$ cut applied on the tagged b quark) forces the contributions
from these two diagrams to be quite small. Thus, even when one is
forced to consider them as part of the NLO QCD corrections to the
$\rm pp\rightarrow p\gamma p\rightarrow pWt$ production process, 
their contribution results for only but a tiny theoretical uncertainty.

Then the cross section for each of the real emission partonic processes
can be written as $\rm \hat{\sigma}^R=\hat{\sigma}^{S}+\hat{\sigma}^{H}=
\hat{\sigma}^{S}+\hat{\sigma}^{HC}+\hat{\sigma}^{\overline{HC}}$ .
After integrating over the photon and quark spectra, we get the real
contributions as $\rm \sigma^{R}=\sigma^S+\sigma^{HC}+\sigma^{\overline{HC}}$.

\subsubsection{Total QCD NLO Cross Section}

After combining all the contributions that are mentioned before, the UV, IR
singularities in our final total cross section
\begin{eqnarray} \nonumber
\rm \sigma^{NLO}&=&\rm \sigma^{LO}+\sigma^{V}+\sigma^{R}\\
            &=&\rm \sigma^{LO}+\sigma^{V}+\sigma^S+\sigma^{HC}+\sigma^{\overline{HC}}
\end{eqnarray}
are exactly cancelled. The logarithmic dependence on the arbitrary small cutoff parameters
$\delta_{s}$ and $\delta_{c}$ are then cancelled (but power-like terms survive).
These cancelations can be
verified numerically in our numerical calculations.
The final results of the total QCD NLO cross section in the 5FNS scheme can be expressed as:
\begin{eqnarray}\label{sigmaNLO}\nonumber
&&\rm \sigma^{NLO} (pp\rightarrow p\gamma p \rightarrow pW^- t+Y)\\\nonumber
&=&\rm  \int^{\sqrt{\xi_{max}}}_{\frac{M_{inv}}{\sqrt{s}}} 2z dz \int^{\xi_{max}}_{Max(z^2,\xi_{min})}
\frac{dx_1}{x_1} f_{\gamma/P_A}(x_1)   \{ G_{b/P_B}(\frac{z^2}{x_1}, \mu_f) [ \hat\sigma^{LO} + \hat\sigma^{V}
+ (F^{soft}+F^{hc}) \hat\sigma^{LO}        \\
&&\rm    + (F^{1}+F^{2}) \hat\sigma^{LO} + \hat\sigma_{g}^{\overline{HC}} ] + G_{g/P_B}(\frac{z^2}{x_1}, \mu_f) \hat\sigma_{\bar{b}}^{\overline{HC}}    \}
   + (A\leftrightarrow B).
\end{eqnarray}
Here $\rm F^{soft}$ and $\rm F^{hc}$ are the factors contain soft
and collinear singularities as well as finite terms. $\rm F^{1,2}$ are
the factors that finite. In the massless b quark assumption, there analytical expression are
\begin{eqnarray}
\rm F^{soft}+F^{hc} &=&\rm  C_F \frac{\alpha_s}{2\pi} (\frac{A_2}{\epsilon^2} +\frac{A_1}{\epsilon} +A_0 )
\end{eqnarray}
with
\begin{eqnarray}\nonumber
\rm  A_2&=&\rm 1 \\\nonumber
\rm  A_1&=&\rm ln\frac{\mu^2_r}{\hat{s}}
     - ln\frac{(e_4-e_{4z})^2}{e_4^2 - e_{4z}^2 - e_{4x}^2} + \frac{5}{2}\\\nonumber
\rm  A_0&=&\rm \frac{1}{2} ln^2\frac{\mu^2_r}{\hat{s}}
     - 2 ln\delta_s ln\frac{\mu^2_r}{\hat{s}}
     + 2 ln^2\delta_s - 2 ln\delta_s + ln\frac{\mu^2_r}{\hat{s}}
     - ln \frac{\mu^2_r}{\hat{s}} ln\frac{(e_4-e_{4z})^2}{e^2_4-e^2_{4z}-e^2_{4x}}  \\ \nonumber
 &+&\rm 2 ln\delta_s ln \frac{(e_4-e_{4z})^2}{e^2_4-e^2_{4z}-e^2_{4x}}
   +
\frac{e_4}{\sqrt{e^2_{4z}+e^2_{4x}}} ln\frac{e_4+\sqrt{e^2_{4z}+e^2_{4x}}}{e_4-\sqrt{e^2_{4z}+e^2_{4x}}} \\\nonumber
 &+&\rm ln^2\frac{e_4-\sqrt{e^2_{4z}+e^2_{4x}}}{e_4-e_{4z}}
 -\frac{1}{2} ln^2 \frac{e_4+\sqrt{e^2_{4z}+e^2_{4x}}}{e_4-\sqrt{e^2_{4z}+e^2_{4x}}} \\
 &+&\rm 2 Li_2\left(\frac{e_{4z}-\sqrt{e^2_{4z}+e^2_{4x}}}{e_4-\sqrt{e^2_{4z}+e^2_{4x}}}\right)
 - 2 Li_2\left(\frac{-e_{4z}-\sqrt{e^2_{4z}+e^2_{4x}}}{e_4-e_{4z}}\right)
\end{eqnarray}
and
\begin{eqnarray}\nonumber
\rm F^{1}&=&\rm \frac{\alpha_s}{2 \pi}  \int^{1-\delta_s}_{\frac{z^2}{x_1}}
  \frac{dy}{y} G_{b/P_B}(\frac{z^2}{x_1 y},\mu_f)
 \left[C_F \frac{1+y^2}{1-y} \rm{ln}\left(\delta_s\frac{1-y}{y}\right)\frac{\hat{s}}{\mu^2_f} +
 \left(1-y\right) \right] \\\nonumber
\rm F^{2}&=&\rm \frac{\alpha_s}{2 \pi} \int^{1}_{\frac{z^2}{x_1}}
 \frac{dy}{y} G_{g/P_B}(\frac{z^2}{x_1 y},\mu_f)
 \left[-\frac{1}{2}(y^2+(1-y)^2) \rm{ln}\left(\delta_s\frac{1-y}{y}\right)\frac{\hat{s}}{\mu^2_f}+y(1-y)\right].
\end{eqnarray}
Notations of $\rm e_{i/ix}$ can be found in Eq.(\ref{Eq.notationE}).
The dilogarithm function $\rm Li_2(x)$ is defined in Ref\cite{Li2}.
Technical details can be found in Ref\cite{2PSS:Owens,2PSS:formula}.

\section{Numerical Results}

\subsection{Input Parameters}

We take the input parameters as $\rm M_p=0.938272046\ GeV$,
$\rm \alpha_{ew}(m^2_Z)^{-1}|_{\overline{MS}}=127.918$,
$\rm m_Z=91.1876 GeV$, $\rm m_W=80.385\ GeV$\cite{2012PDG}
and we have $\rm sin^2\theta_W=1-(m_W/m_Z)^2=0.222897$.
The PDFs are taken from the LHAPDF package\cite{LHAPDF}.
We adopt the CTEQ6L1 and CTEQ6M PDFs\cite{CTEQ6} for the LO 
and QCD higher order calculations, separately. 
The strong coupling constant $\rm \alpha_s(\mu)$ is determined by the QCD parameter 
$\rm \Lambda^{LO}_5 = 165 MeV$ for the CTEQ6L1 and 
$\rm \Lambda^{\overline{MS}}_5 = 226 MeV$ for the CTEQ6M, respectively.
For simplicity we set the factorization scale and the renormalization scale
being equal (i.e., $\rm \mu=\mu_f=\mu_r$) and take
$\rm \mu=\mu_0=(m_t+m_W)/2$ in default unless otherwise stated.
Throughout this paper, we set the quark masses as $\rm m_u=m_d=m_c=m_s=m_b=0$,
The top quark pole mass is set to be $\rm m_t=173.5\ GeV$.
The colliding energy in the proton-proton center-of-mass system
is assumed to be $\rm \sqrt{s}=14\ TeV$ at future LHC.
We adopt BASES\cite{BASES} to do the phase space integration.
The CKM matrix elements are set as unit.
The decay of the top quark is expected to be dominated
by the two-body channel $\rm t\rightarrow W^- b$
and the total decay width of the top quark
is approximately equal to the decay width of $\rm t\rightarrow W^- b$.
Neglecting terms of order $\rm m^2_b/m^2_t$, $\rm \alpha_s$,
and $\rm (\alpha_s/\pi) m^2_W/m^2_t$, the width predicted in the SM at NLO is:
\begin{eqnarray}
\rm \Gamma_t =\rm \frac{\alpha_{ew}m^3_t}{16 m^2_W s^2_W}
 \left(1-\frac{m^2_W}{m^2_t}\right)^2 \left(1+\frac{2 m^2_W}{m^2_t}\right)
 \left[ 1-\frac{2\alpha_s}{3\pi} \left(\frac{2\pi^2}{3}-\frac{5}{2}\right) \right].
\end{eqnarray}
By taking  $\rm \alpha_{ew}(m^2_Z)^{-1}|_{\overline{MS}}=127.918$
and $\rm \alpha_s(m^2_t)=0.1079$, we obtain $\rm \Gamma_t=1.41595\ GeV$.
Based on the forward proton detectors to be installed by the CMS-TOTEM and
the ATLAS collaborations we choose the detected acceptances to be
\begin{itemize}
 \item CMS-TOTEM forward detectors with $0.0015<\xi_1<0.5$
 \item CMS-TOTEM forward detectors with $0.1<\xi_2<0.5$
 \item AFP-ATLAS forward detectors with $0.0015<\xi_3<0.15$
\end{itemize}
which we simply refer to $\xi_1$, $\xi_2$ and $\xi_3$, respectively.
During calculation we use $\xi_1$ in default unless otherwise stated.
Note here we do not consider the decay of the heavy final states
as well as the survival probability in the $\rm \gamma p$ collision
or simply taken to be unit.

\subsection{Cross Checks}

Before presenting the numerical predictions, several cross checks should be done.

\begin{itemize}
\item First, during the calculation of the tensor one loop integrals,
we use our modified LoopTools and cross check with OneLoop\cite{Oneloop} and QCDLoop\cite{QCDloop} packages.
We can get exactly the same results in each phase space point.

\item Second, when do the phase space integration we use BASES\cite{BASES} and cross check
independently with Kaleu\cite{Kaleu} especially for the hard emission contributions.
We can get the same integrated results within the error.

\item Third, the UV and IR safeties should be verified numerically after combining all
the contributions at the NLO QCD loop level. To check this, we display enough random phase
space point as well as the cancellation for different divergent parameters, see in Table.\ref{tab1}
corresponding to 5FNS massless b quark scheme.
One thing that should be emphasized is $\rm \hat\sigma^{V}$ should include the counter-term contributions
as well as the soft and collinear singularity terms coming from the real emissions.
We implement this into our monte carlo codes which provide an automatic check of the dependence
on these divergence parameters. We can see the UV and IR divergence can be canceled at high precision
level in all the phase space thus leading the continuance of our following calculation.

\begin{table}[hbtp]
\begin{center}
\begin{tabular}{c c c}
\hline
\hline
 \multicolumn{3}{c}{
\hspace*{-1.2cm} $\rm pp\rightarrow p\gamma p\rightarrow p\gamma(p_1) b(p_2)\rightarrow p W(p_3)t(p_4)+Y$ with $\rm m_b=0$ [pb]} \\
\hline
 \multicolumn{3}{c}{\hspace*{-3.5cm}$\rm p_1$=(218.59020657143321 0 0 218.59020657143321)}\\
 \multicolumn{3}{c}{\hspace*{-3.4cm}$\rm p_2$=(218.59020657143321 0 0 -218.59020657143321)}\\
 \multicolumn{3}{c}{\hspace*{-0.1cm}$\rm p_3$=(191.55273981365957 148.54834482485680 0 90.355371477931513)}\\
 \multicolumn{3}{c}{\hspace*{0.1cm}$\rm p_4$=(245.62767332920686 -148.54834482485680 0 -90.355371477931513)}\\
\hline
$\hspace*{-0.5cm}\rm \frac{1}{\epsilon_{UV}}=\frac{1}{\epsilon^2_{IR}}=\frac{1}{\epsilon_{IR}}=0$
&$\rm \hat\sigma^{LO}=0.7990$ & $\rm \hat\sigma^{V}=\hat\sigma^0=2.2672370626972782   $  \\
$\rm \frac{1}{\epsilon_{UV}}=\frac{1}{\epsilon^2_{IR}}=\frac{1}{\epsilon_{IR}}=10^{10}$
&$\rm \hat\sigma^{LO}=0.7990$ &\hspace*{-2.1cm}  $\rm \hat\sigma^{V}=\hat\sigma^0+7\times 10^{-17}$  \\
\hline
\hline
\end{tabular}
\end{center}
\vspace*{-0.8cm}
\begin{center}
\begin{minipage}{14cm}
\caption{\label{tab1} The UV and IR divergence cancelation at one given
random phase space point for the loop contribution for
$\rm pp\rightarrow p\gamma p\rightarrow p\gamma(p_1) b(p_2)\rightarrow p W(p_3)t(p_4)+Y$ with $\rm m_b=0$. }
\end{minipage}
\end{center}
\end{table}

\item Fourth, since the total cross section is independent of the soft cutoff
$\rm \delta_s$ $\rm(=\Delta E_g/E_b, E_b=\sqrt{\hat{s}}/2)$ and the collinear cutoff
$\rm \delta_c$, we display their values for $\rm pp\rightarrow p\gamma p\rightarrow pWt$
versus the cutoff $\rm \delta_s$, where we take $\rm \delta_c=\delta_s/100$.
Both $\rm \delta_s$ and $\rm \delta_c$ dependence should be checked.
Some of the results are listed in Table.\ref{tab2}.
The detector acceptance here is chosen to be $0.0015<\xi_1<0.5$.
It is shown clearly that the NLO QCD correction does not depend on the arbitrarily
chosen values of $\rm \delta_s$ and $\rm \delta_c$ within the calculation errors.
In the further numerical calculations, we fix $\rm \delta_s = 10^{-4}$ and $\rm \delta_c=\delta_s/100$.

\begin{table}
\begin{center}
\begin{tabular}{c c c c c c}
\hline\hline
\multicolumn{6}{c}{$\rm \delta_s$ dependence for $\rm \sigma(pp\rightarrow p\gamma p\rightarrow pWt+Y)$[pb]with $\rm m_b=0$} \\ [0.5ex]
 $\rm \delta_s=100\delta_c$ && $\rm \sigma^{LO}$ && $\rm \sigma^{NLO}$ &$\rm \Delta=\sigma^{NLO}-\sigma^{LO}$  \\
\hline
$1\times 10^{-3.0}$ &&1.14255073 &&1.34824330 & 0.20569257\\
$1\times 10^{-3.5}$ &&1.14255073 &&1.34968444 & 0.20713371\\
$1\times 10^{-4.0}$ &&1.14255073 &&1.34910699 & 0.20655626\\
$1\times 10^{-4.5}$ &&1.14255073 &&1.34921708 & 0.20666635\\
$1\times 10^{-5.0}$ &&1.14255073 &&1.34937836 & 0.20682763\\
$1\times 10^{-5.5}$ &&1.14255073 &&1.35015998 & 0.20760925\\
$1\times 10^{-6.0}$ &&1.14255073 &&1.34720592 & 0.20465519\\
\hline\hline
\end{tabular}
\end{center}
\vspace*{-0.8cm}
\begin{center}
\begin{minipage}{14cm}
\caption{\label{tab2}
The $\rm \delta_s$ dependence of the loop induced QCD correction to the integrated
cross section for the $\rm pp\rightarrow p\gamma p \rightarrow pWt+Y$
with $\rm m_b=0$
at the $\rm \sqrt{s}=14\ TeV$ LHC where we set $\rm \delta_c=\delta_s/100$. The detector
acceptance here is chosen to be $0.0015<\xi_1<0.5$.}
\end{minipage}
\end{center}
\end{table}

\end{itemize}

\subsection{Scale dependence for different forward detector acceptances}

\begin{figure}[hbtp]
\centering
\includegraphics[scale=0.6]{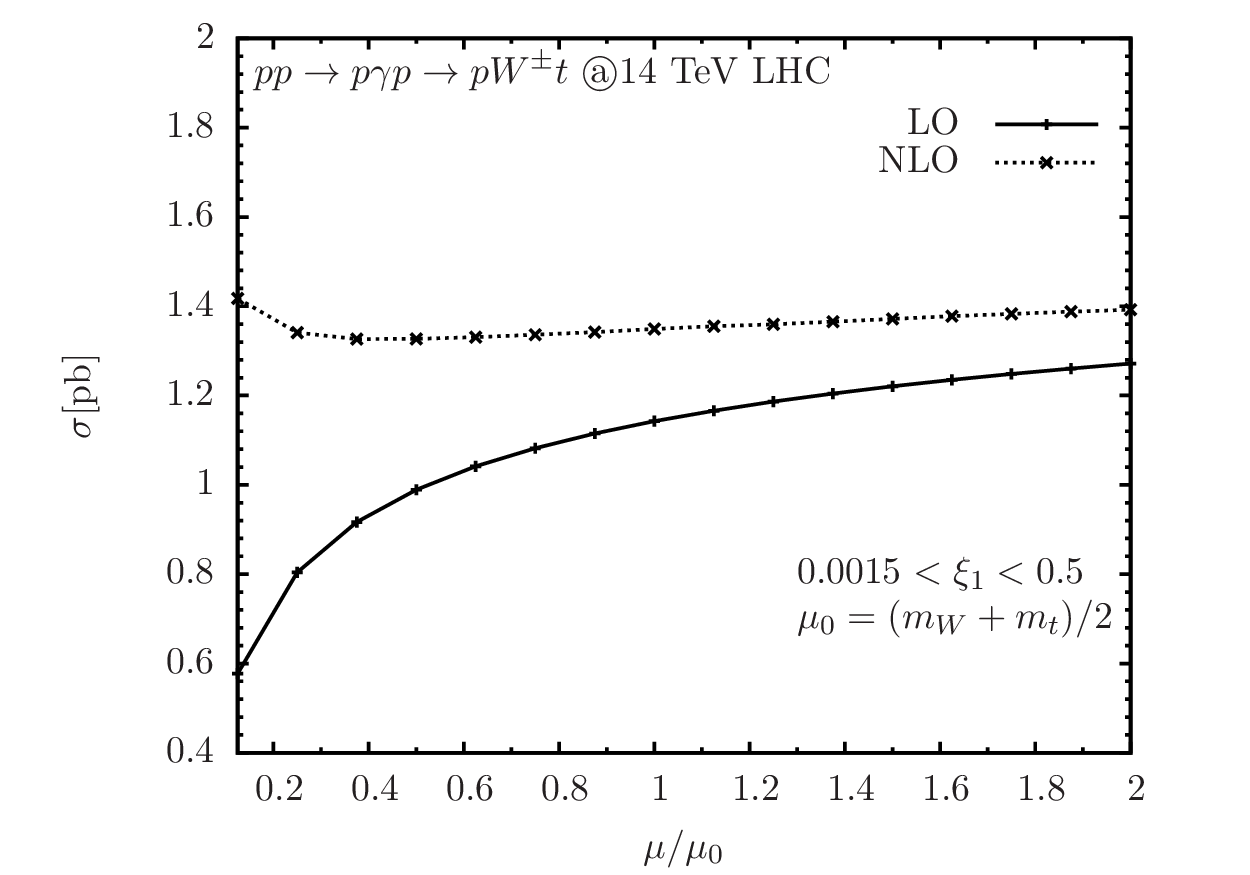}
\includegraphics[scale=0.6]{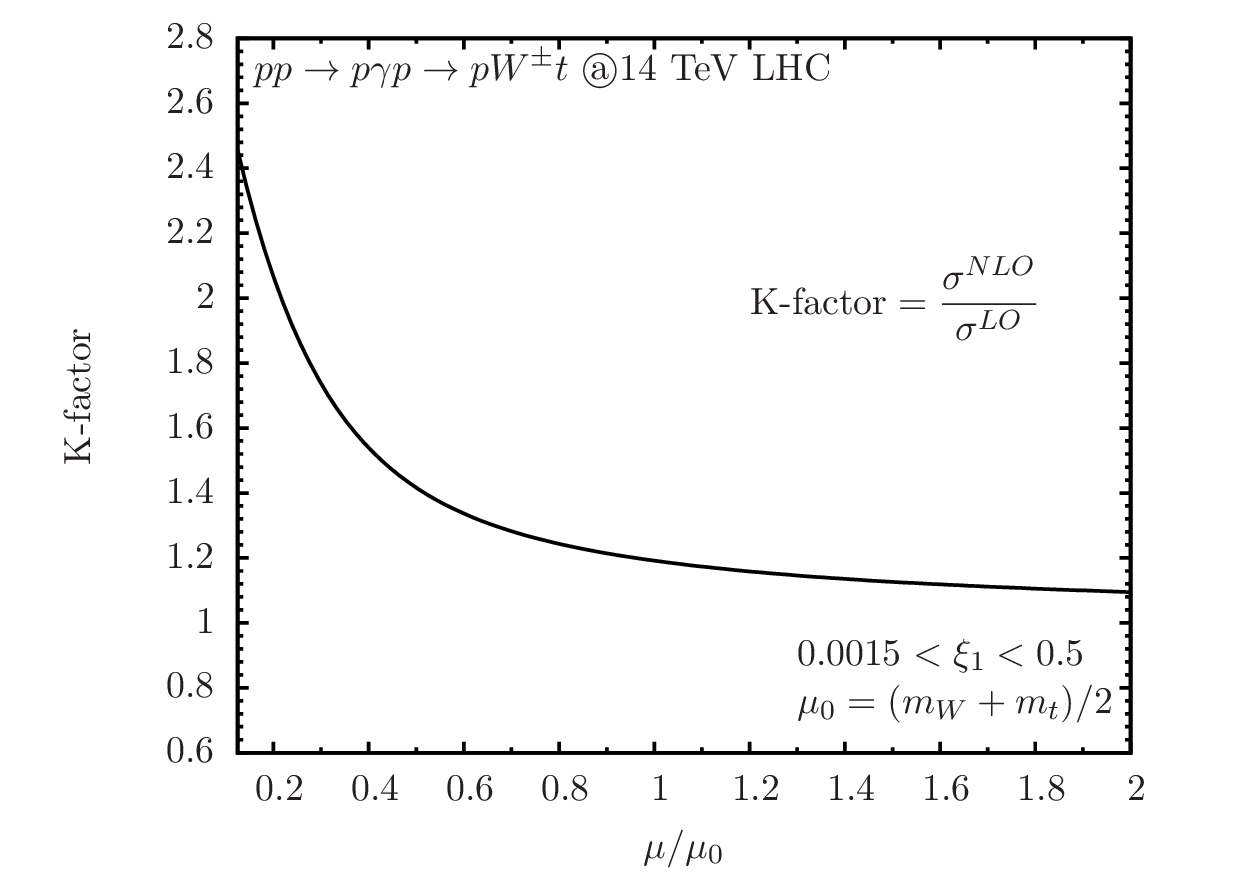}
\caption{\label{fig6}
The scale-$\mu$ dependence of the LO and QCD NLO corrected cross sections  [left panel]
and K-factor [right panel] for $\rm pp\rightarrow p\gamma p \rightarrow pWt+Y$ at the
$\rm \sqrt{s}= 14\ TeV$ LHC with $\rm \mu_0=(m_W+m_t)/2$, $\rm \delta_s=10^{-4}$
and $\rm \delta_c=\delta_s/100$. The experimental acceptance here
is chosen to be $0.0015<\xi_1<0.5$ for the CMS-TOTEM forward detectors.}
\end{figure}

We present the scale-$\mu$ dependence of the LO and QCD NLO
corrected cross sections for $\rm pp\rightarrow p\gamma p\rightarrow pWt+Y$
for the CMS-TOTEM forward detectors with $0.0015 < \xi_1< 0.5$
in the left panel of Fig.\ref{fig6}. Scale-$\mu$ varies from $\mu_0/8$
to $2\mu_0$ with $\rm \mu_0=(m_W+m_t)/2$.
In the figure, solid lines with plus sign points present the LO predictions.
Its cross section varies from 0.5772 pb to 1.2717 pb with
the scale-$\mu$ varies from $\mu_0/8$ to $2\mu_0$. The deviation
is as large as 0.6945 pb shows some dependence on the scale.
We use dotted line with times sign to present the QCD NLO corrected cross section
in the 5FNS massless b quark scheme. The NLO cross section changes
from 1.4175 pb to 1.3922 pb with the deviation only 0.0253 pb.
We can see that if the QCD NLO corrections are taken into account,
much better scale-$\mu$ independence can be obtained and
the factorization/renormalization scale uncertainty can be reduced.
In the right panel of Fig.\ref{fig6}, we show the
K-factor of the QCD NLO contribution as function of scale-$\mu$.
K-factor is defined as $\rm \sigma^{NLO}/\sigma^{LO}$.
We see that K-factor is large and sensitive in the small $\mu$ range while
insensitive at the large $\mu$.
Typical results of the K-factor are
2.4557, 1.3415, 1.1808 and 1.0947 for $\mu_0/8$, $\mu_0/2$,
$\mu_0$ and $2\mu_0$, respectively.

\begin{figure}[hbtp]
\centering
\includegraphics[scale=0.6]{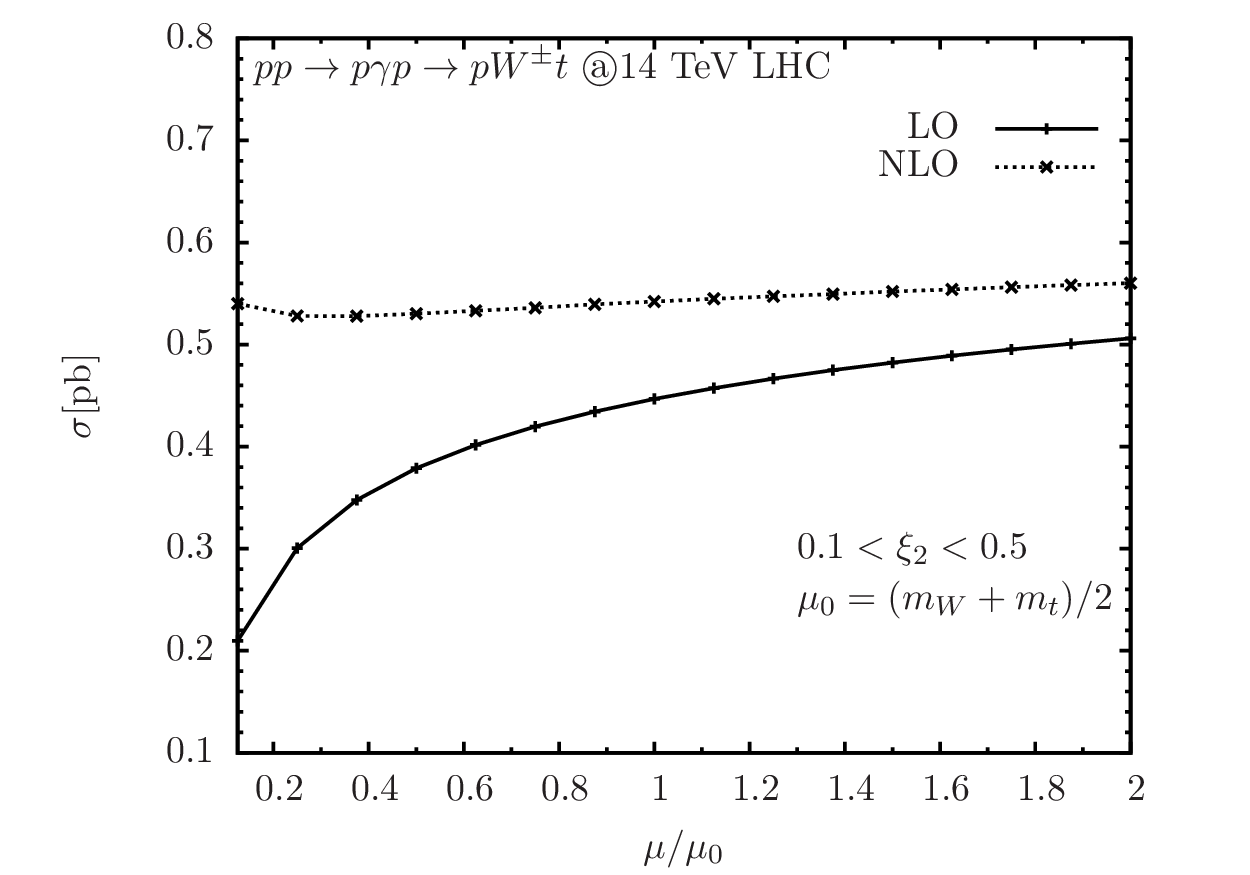}
\includegraphics[scale=0.6]{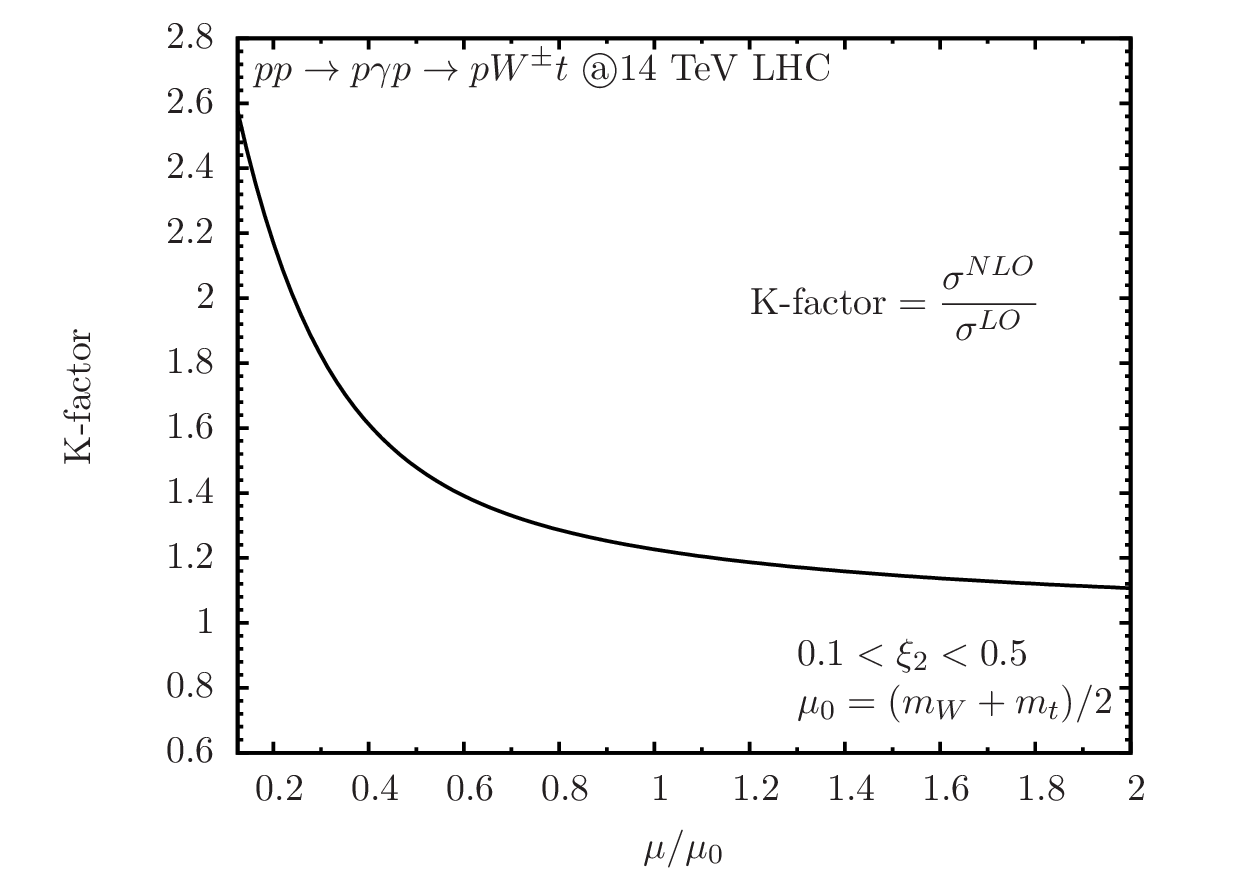}
\caption{\label{fig7}
The scale-$\mu$ dependence of the LO and QCD NLO corrected cross sections  [left panel]
and K-factor [right panel] for $\rm pp\rightarrow p\gamma p \rightarrow pWt+Y$ at the
$\rm \sqrt{s}= 14\ TeV$ LHC with $\rm \mu_0=(m_W+m_t)/2$, $\rm \delta_s=10^{-4}$
and $\rm \delta_c=\delta_s/100$. The experimental acceptance here
is chosen to be $0.1<\xi_2<0.5$ for the CMS-TOTEM forward detectors.}
\end{figure}

In Fig.\ref{fig7}, the scale-$\mu$ dependence of the LO cross section,
QCD NLO corrected cross section and K-factor are depicted in the left
and right panel for the CMS-TOTEM forward detectors with $0.1<\xi_2<0.5$.
Same as in Fig.\ref{fig6}, we use solid lines with plus sign points present
the LO predictions and dotted line with times sign to present
the QCD NLO corrected cross section, respectively.
When $\mu$ varies from $\mu_0/8$ to $2\mu_0$, their cross sections
change from 0.2097(0.5485) pb to 0.5401(0.5601) pb with their ratio equal 2.58 (1.02).
Still we can see the NLO predictions can reduce the
factorization/renormalization scale uncertainty corresponding to the LO prediction.
We see NLO correction shows much better scale
independence through the whole range $[\mu_0/8,2\mu_0]$.
Compare the results with $0.0015 < \xi_1< 0.5$, we see for $0.1<\xi_2<0.5$,
in the whole range, both the LO and QCD NLO corrected cross sections
are smaller, less than half of than that of $0.0015 < \xi_1< 0.5$.

\begin{figure}[hbtp]
\centering
\includegraphics[scale=0.6]{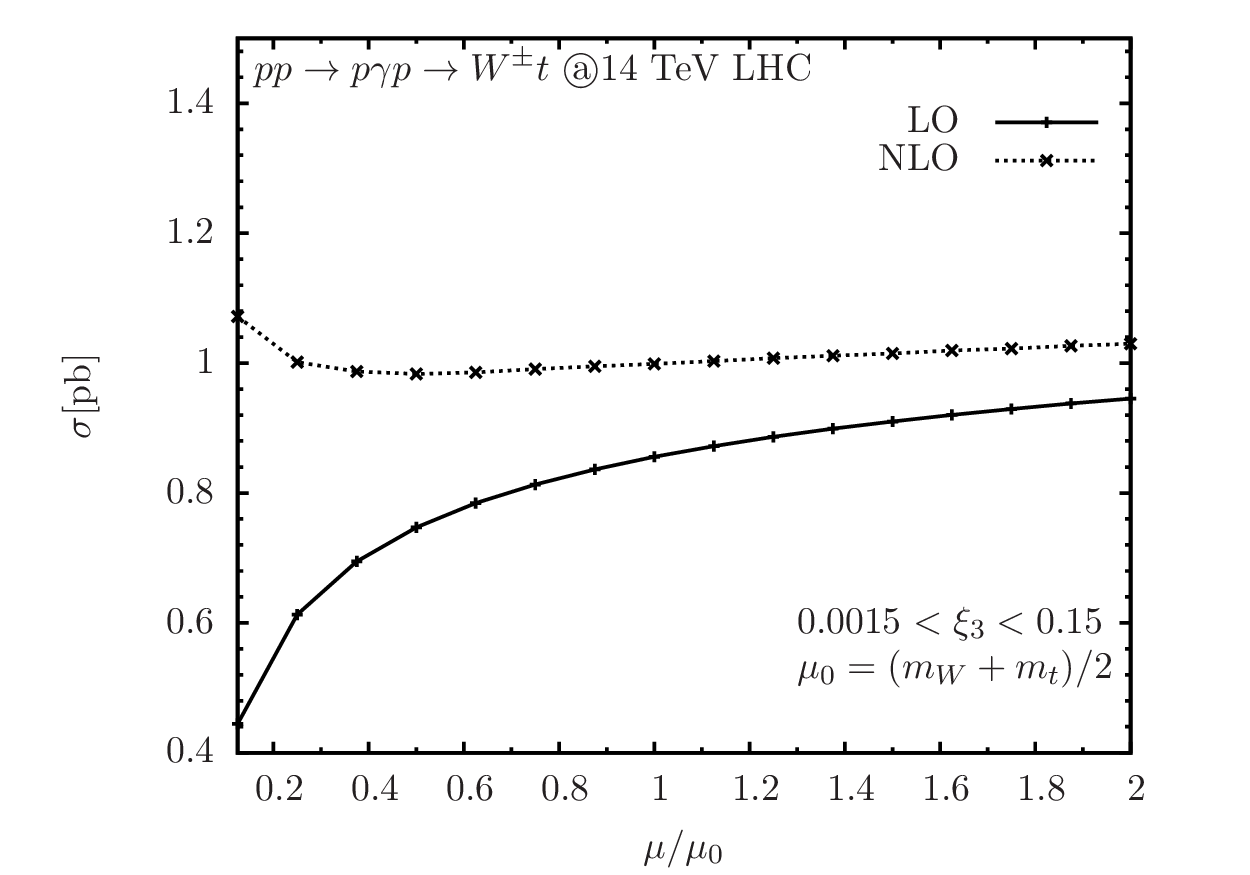}
\includegraphics[scale=0.6]{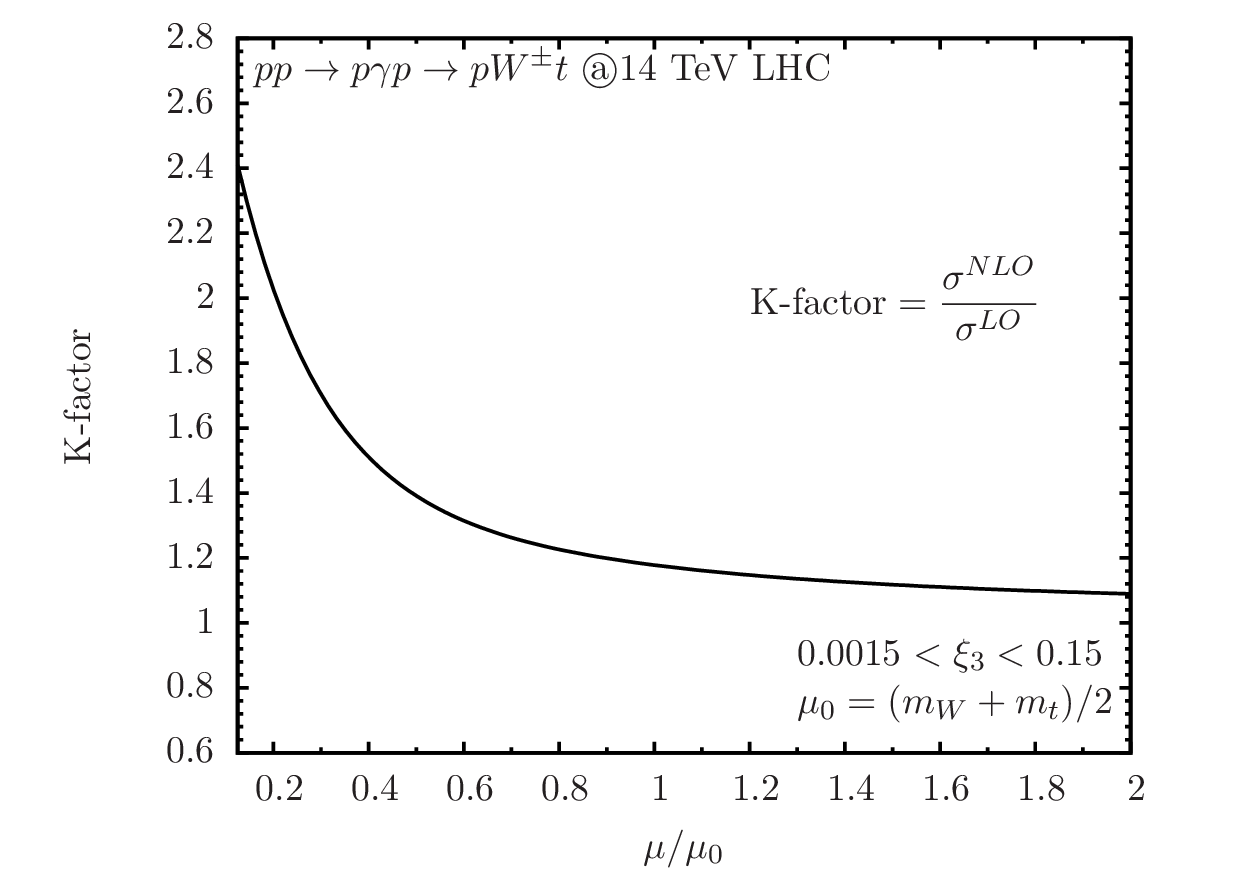}
\caption{\label{fig8}
The scale-$\mu$ dependence of the LO and QCD NLO corrected cross sections  [left panel]
and K-factor [right panel] for $\rm pp\rightarrow p\gamma p \rightarrow pWt+Y$ at the
$\rm \sqrt{s}= 14\ TeV$ LHC with $\rm \mu_0=(m_W+m_t)/2$, $\rm \delta_s=10^{-4}$
and $\rm \delta_c=\delta_s/100$. The experimental acceptance here
is chosen to be $0.0015<\xi_3<0.15$ for AFP-ATLAS forward detectors.}
\end{figure}

For the AFP-ATLAS forward detectors with $0.0015<\xi_3<0.15$,
the cross section for the LO and QCD NLO prediction are close to that of $0.0015<\xi_1<0.5$,
see Fig.\ref{fig8}.
Behavior of the cross sections and K-factor on the scale-$\mu$ dependence are the same.
In this case their cross sections change from 0.4447(1.0717) pb
to 0.9454(1.0297) pb with their ratio equal 2.13 (0.96)
when $\mu$ varies from $\mu_0/8$ to $2\mu_0$.
We conclude again that the QCD NLO corrections can reduce the factorization/renormalization
scale uncertainty.
Finally we summary the K-factor
for typical value of $\mu$ in Table.\ref{tab3} for different forward detector acceptances.
In our further calculations we fix $\rm \mu=\mu_0=(m_W+m_t)/2$.

\begin{table}
\begin{center}
\begin{tabular}{c c c c c c}
\hline\hline
\multicolumn{6}{c}{K-factor for $\rm pp\rightarrow p\gamma p \rightarrow pWt+Y$}   \\
\hline
$ \xi \backslash \mu$ && $\mu_0/8$ & $\mu_0/2$ & $\mu_0$  & $2\mu_0$\\
\hline
$\xi_1$ &&2.4557    & 1.3415     &  1.1808  & 1.0947 \\
$\xi_2$ &&2.5761     & 1.3997    &  1.2139  &1.1069 \\
$\xi_3$ &&2.4101     & 1.3166    &  1.1673  & 1.0892 \\
\hline\hline
\end{tabular}
\end{center}
\vspace*{-0.8cm}
\begin{center}
\begin{minipage}{14cm}
\caption{\label{tab3}
The K-factor for typical value of $\mu$ for different forward detector acceptances
$0.0015<\xi_1<0.5$, $0.1<\xi_2<0.5$ and $0.0015<\xi_3<0.15$ with $\rm \mu_0=(m_W+m_t)/2$.}
\end{minipage}
\end{center}
\end{table}

\subsection{Distribution and Cross Section}

\begin{figure}[hbtp]
\begin{centering}

\subfloat[$0.0015<\xi_1<0.5$]
{\begin{centering}
\includegraphics[scale=0.6]{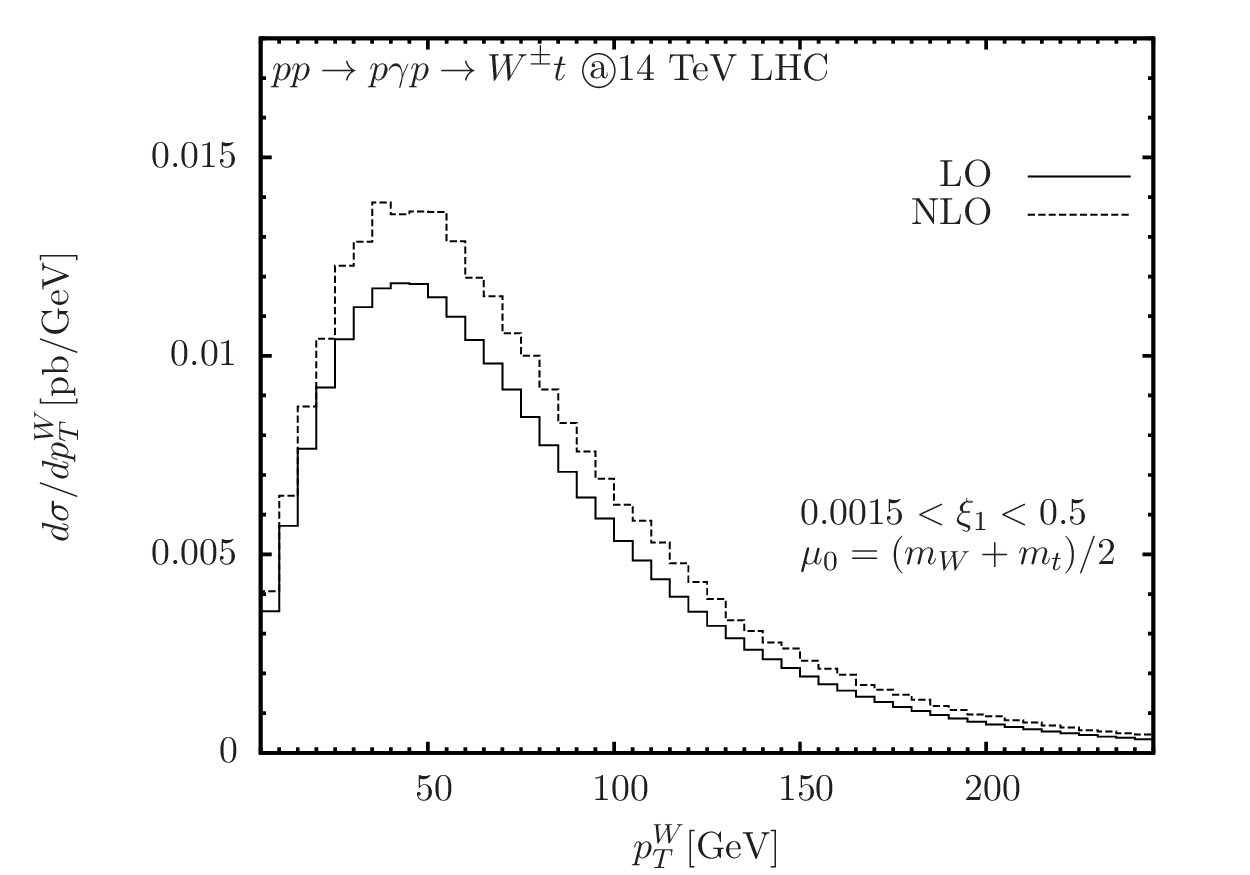}
\includegraphics[scale=0.6]{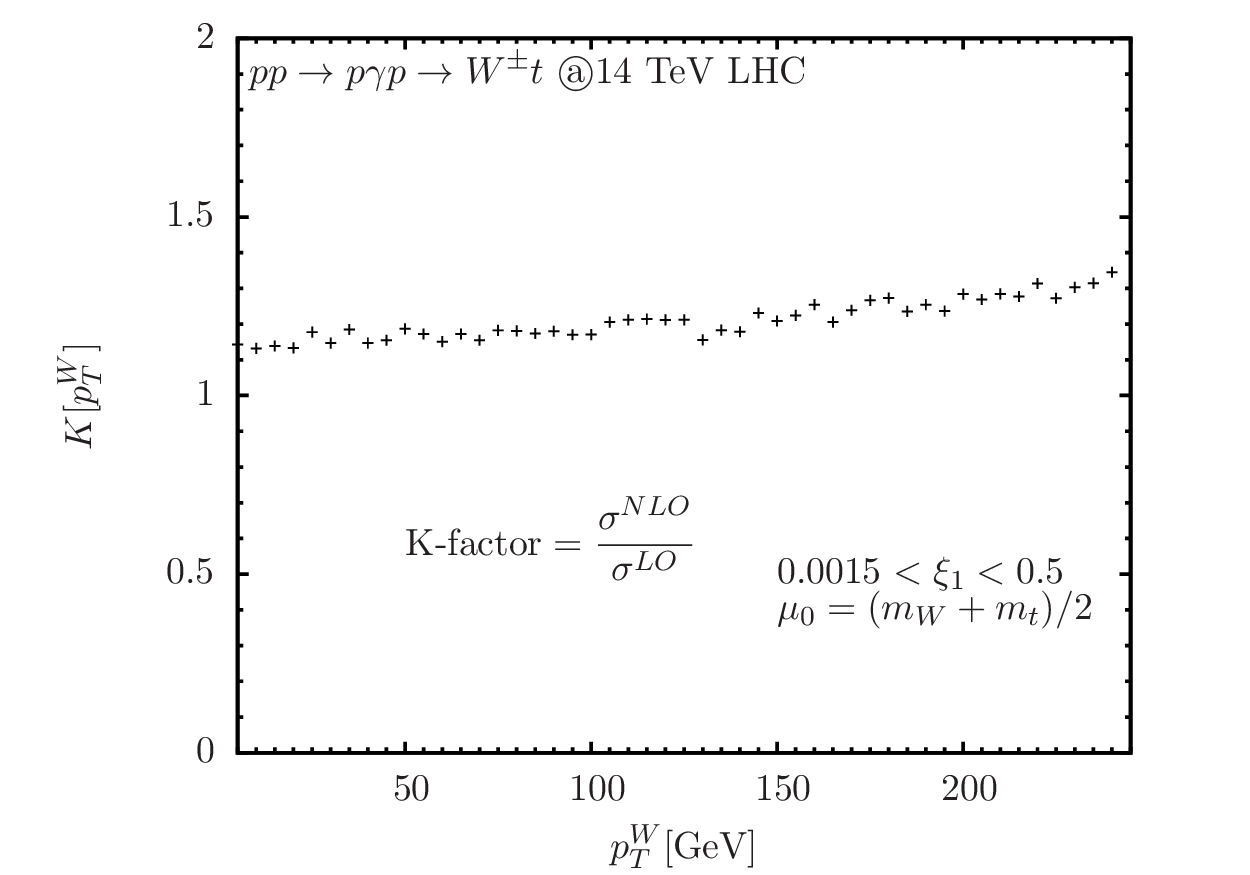}
\par\end{centering}}

\subfloat[$0.1<\xi_2<0.5$]
{\begin{centering}
\includegraphics[scale=0.6]{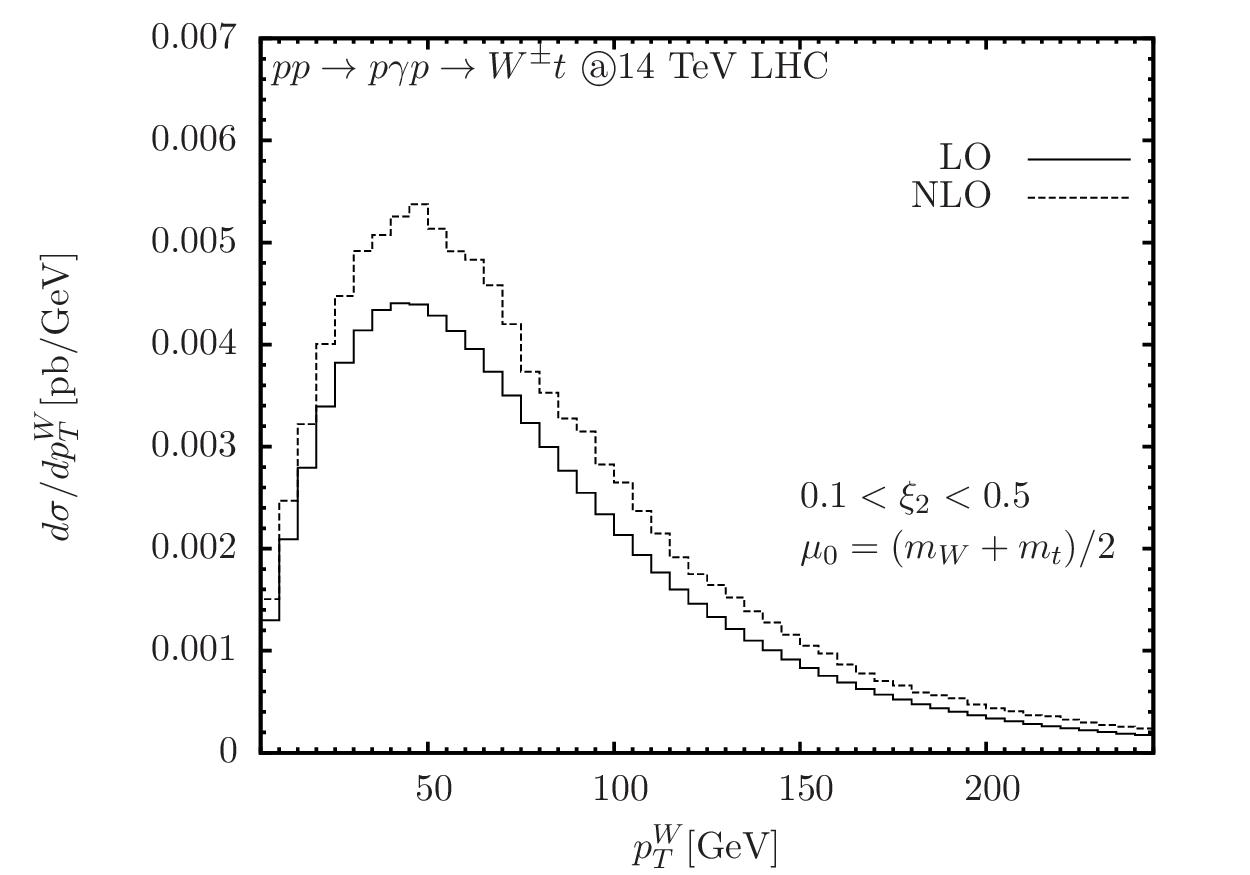}
\includegraphics[scale=0.6]{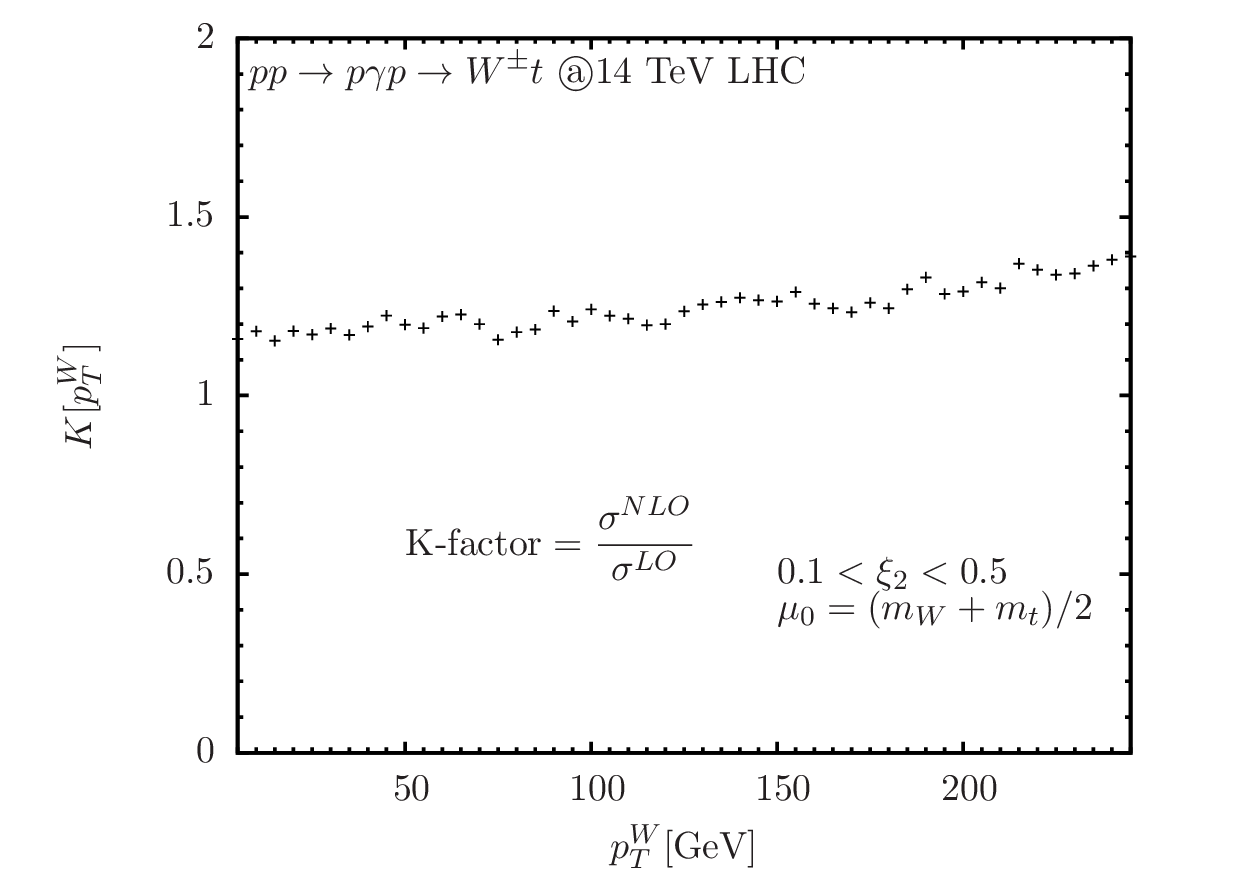}
\par
\end{centering}}

\subfloat[$0.0015<\xi_3<0.15$]{\begin{centering}
\includegraphics[scale=0.6]{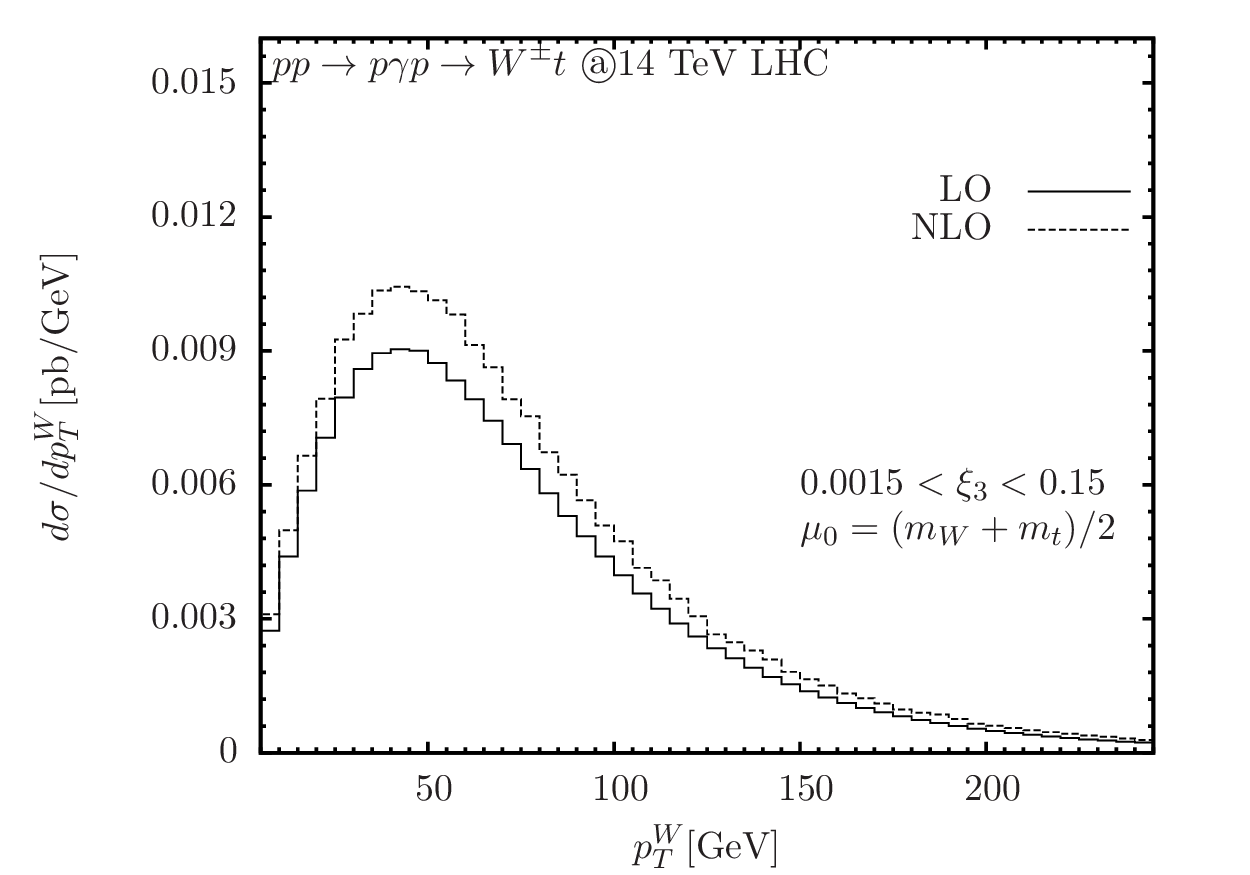}
\includegraphics[scale=0.6]{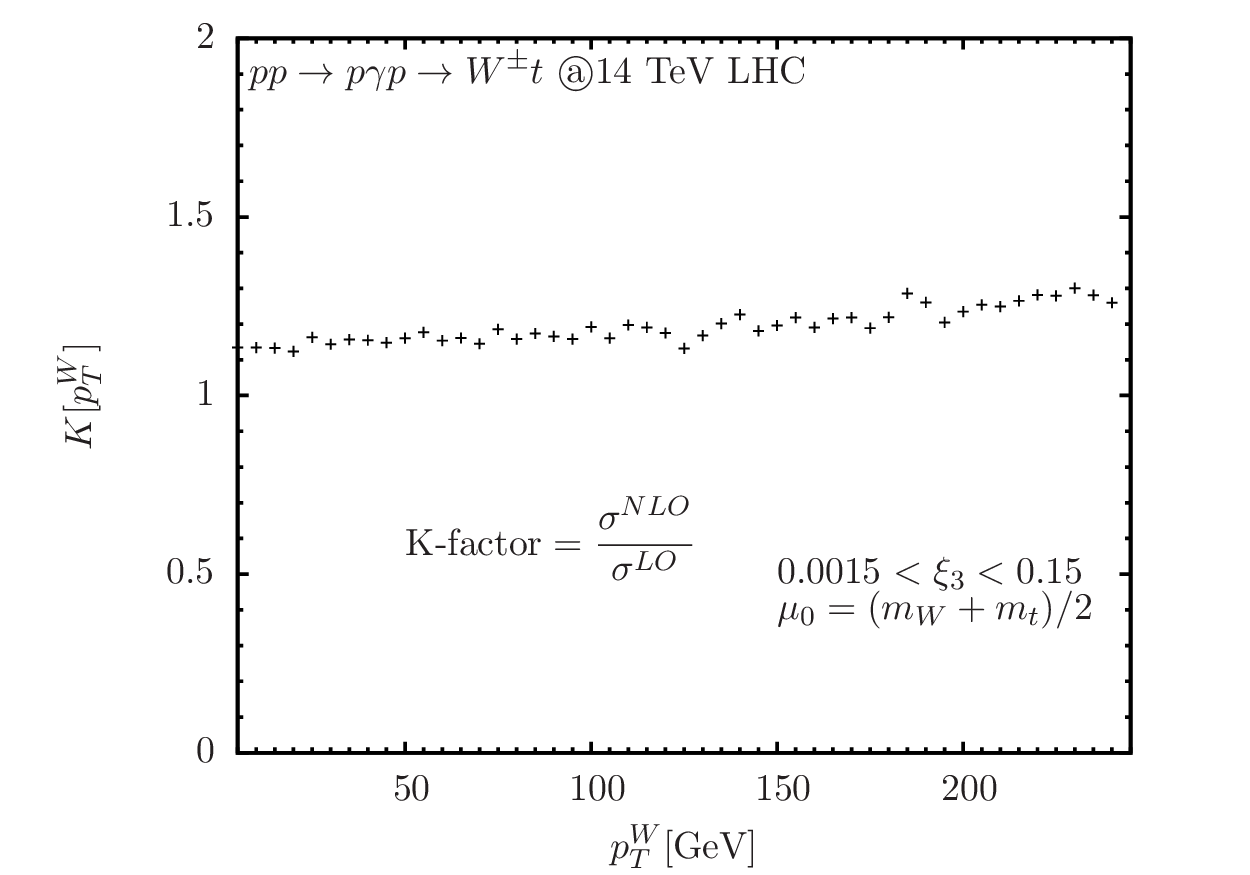}
\par
\end{centering}}

\par
\end{centering}
 \caption{\label{fig9} The LO (solid curves) and NLO (dotted curves) transverse momentum ($p_T$)
distribution of W [left panel] and K-factor [right panel]
for the process $\rm pp\rightarrow p\gamma p\rightarrow pWt+Y$
at 14 TeV LHC with $\rm \mu=\mu_0=(m_W+m_t)/2$, $\rm \delta_s=10^{-4}$ and $\rm \delta_c=\delta_s/100$.
The experimental detector acceptances are $0.0015<\xi_1<0.5$(a), $0.1<\xi_2<0.5$(b), $0.0015<\xi_3<0.15$(c).}
\end{figure}

In Fig.\ref{fig9}, we show the LO (solid curves) and NLO (dotted curves)
transverse momentum ($\rm p_T$) distribution of W [left panel] and K-factor [right panel]
for the process $\rm pp\rightarrow p\gamma p\rightarrow pWt+Y$.
(a), (b) and (c) for the experimental detector acceptances $\xi_1$, $\xi_2$ and $\xi_3$, respectively.
Of course the distributions depend on the detector acceptances,
i.e., in the most efficient case $0.0015<\xi_1<0.5$. However,
their line behaviors are the same for different detector acceptances.
For the distributions of $\rm p_T^{top}$, their behaviors are very similar to those of
$\rm p_T^{W}$ thus not shown. The results show that
QCD NLO contribution can enhance the LO distribution in the whole $\rm p_T$ range.
Typical K-factor is in the range of [1.1-1.4].

\begin{figure}[hbtp]
\begin{centering}
\subfloat[$0.0015<\xi_1<0.5$]
{\begin{centering}
\includegraphics[scale=0.6]{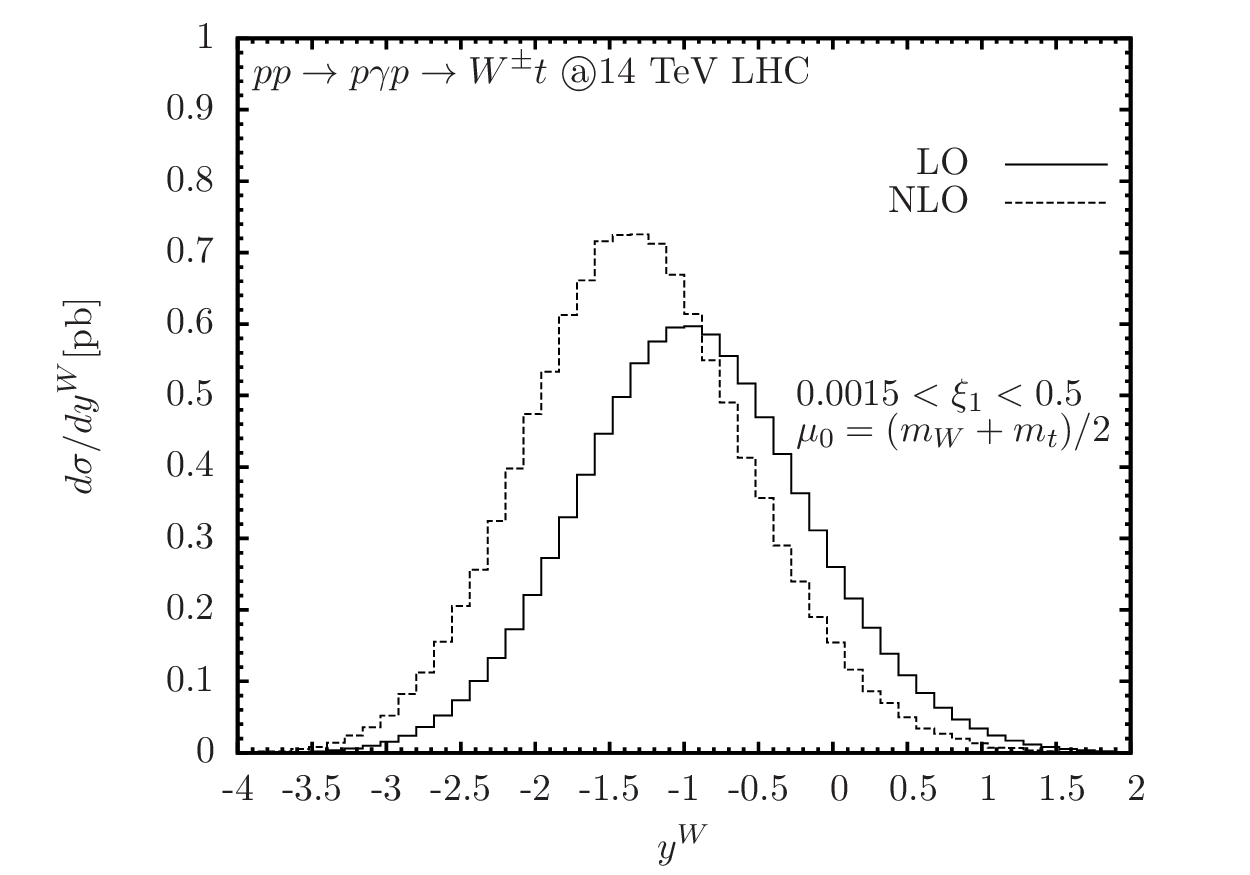}
\includegraphics[scale=0.6]{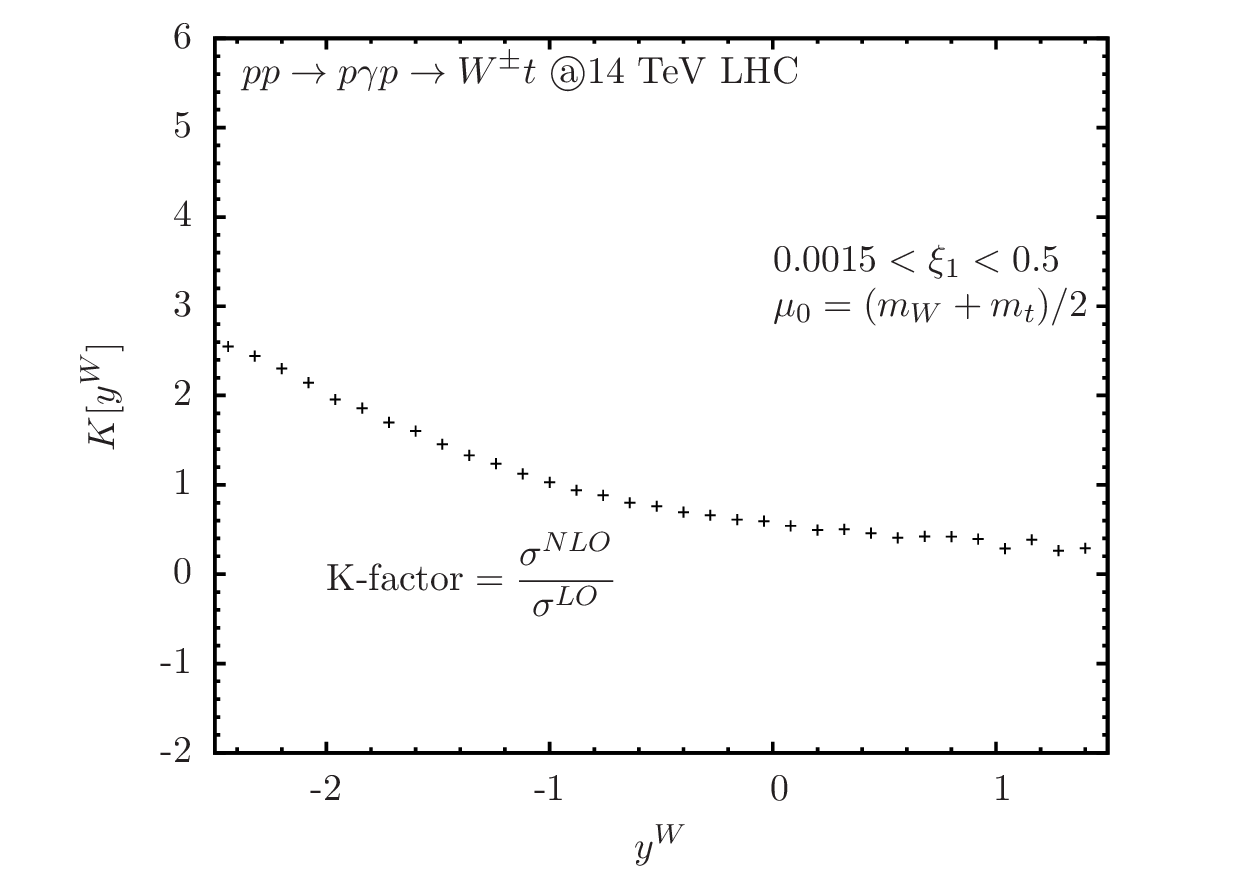}
\par\end{centering}}

\subfloat[$0.1<\xi_2<0.5$]
{\begin{centering}
\includegraphics[scale=0.6]{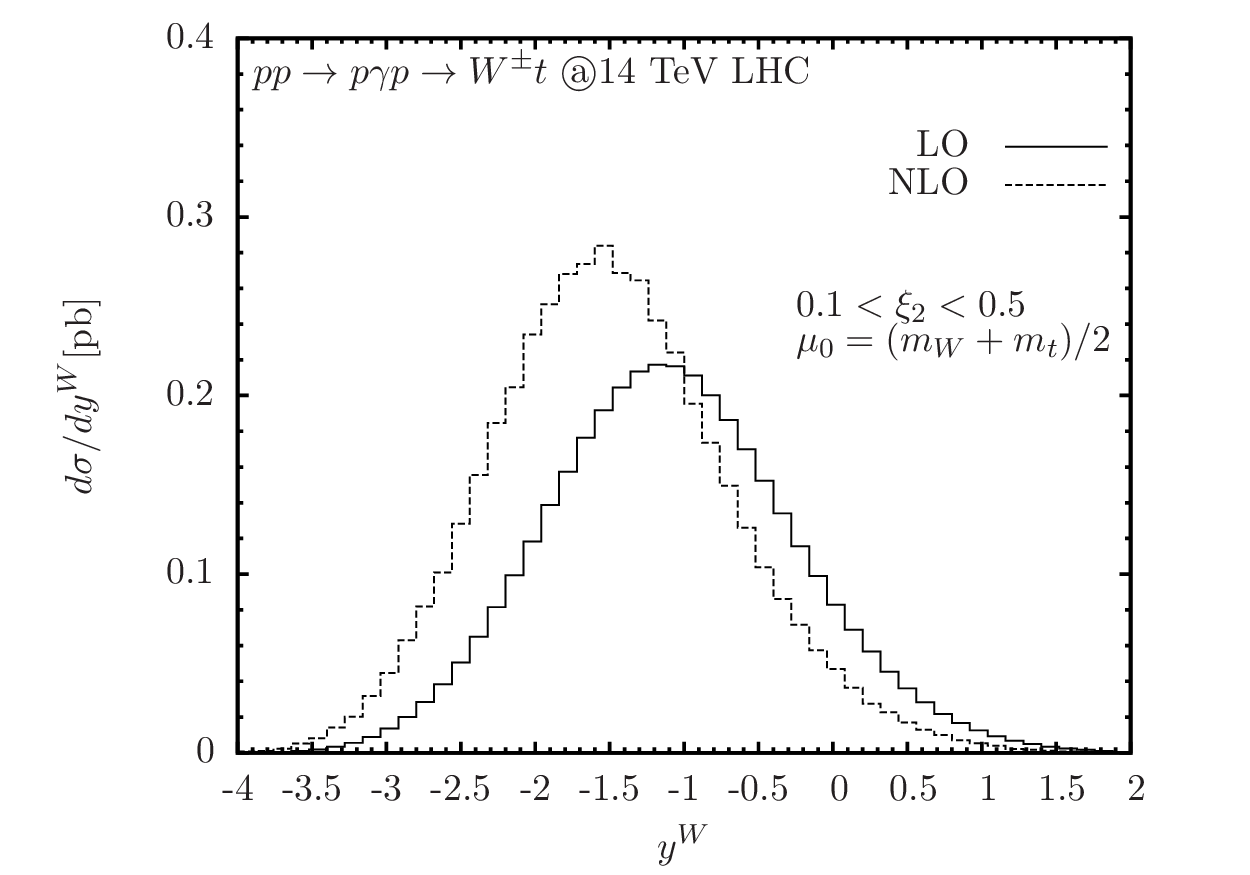}
\includegraphics[scale=0.6]{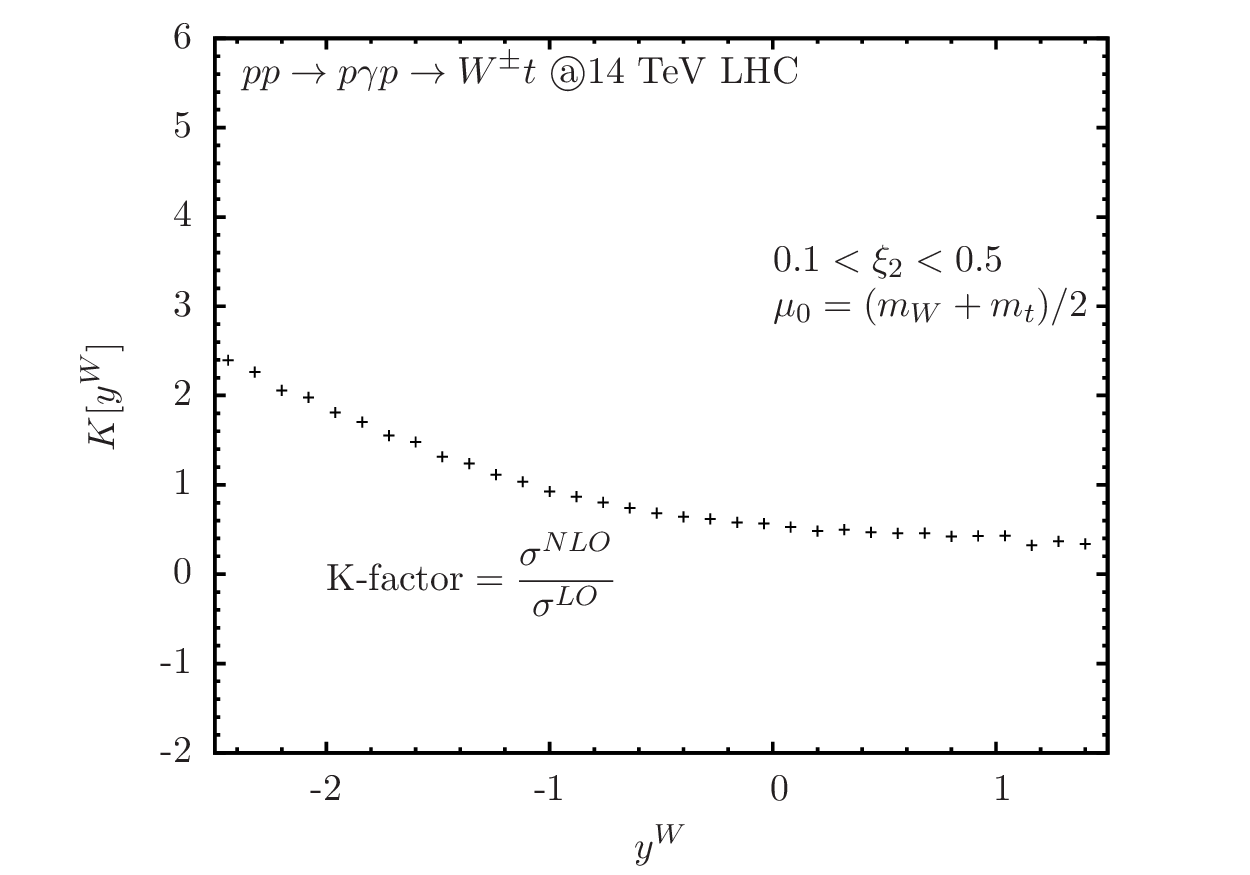}
\par
\end{centering}}

\subfloat[$0.0015<\xi_3<0.15$]
{\begin{centering}
\includegraphics[scale=0.6]{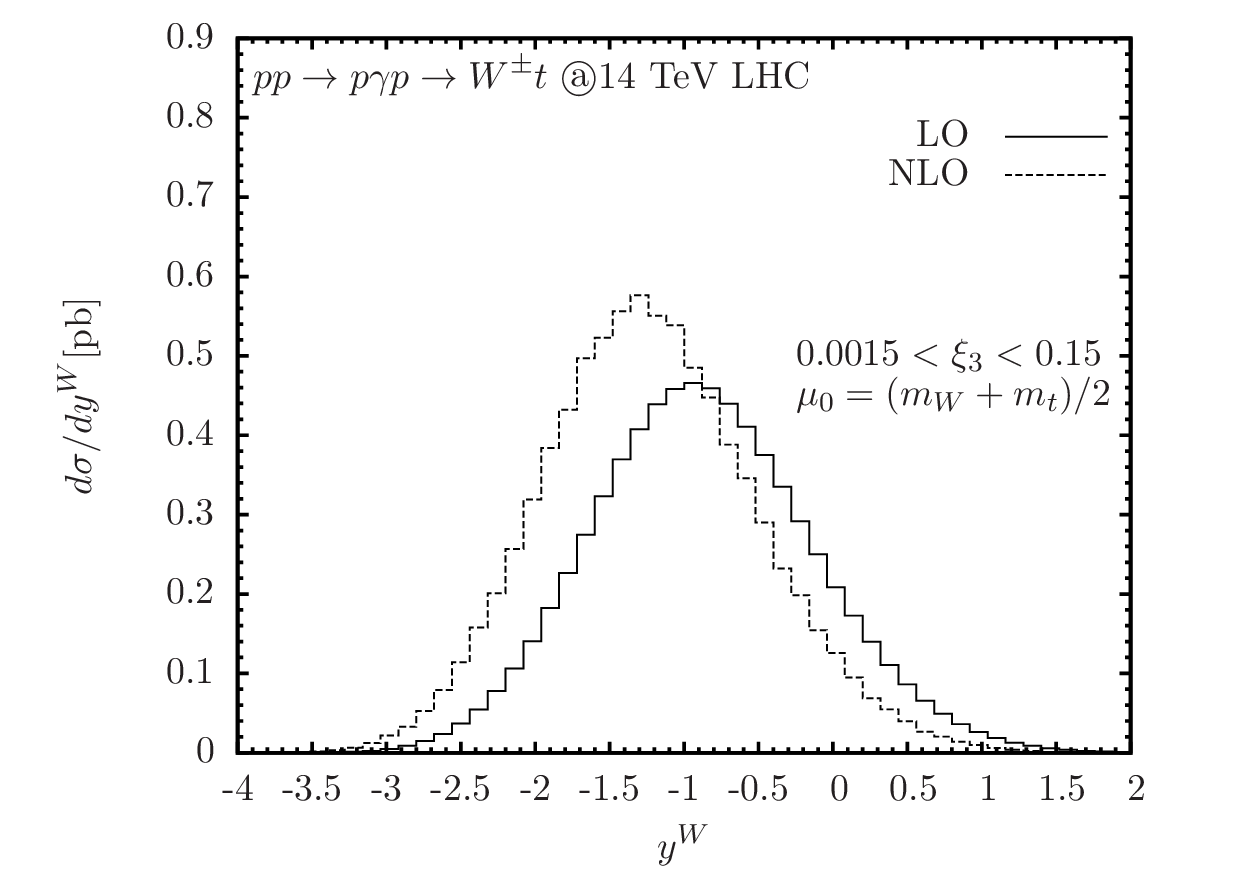}
\includegraphics[scale=0.6]{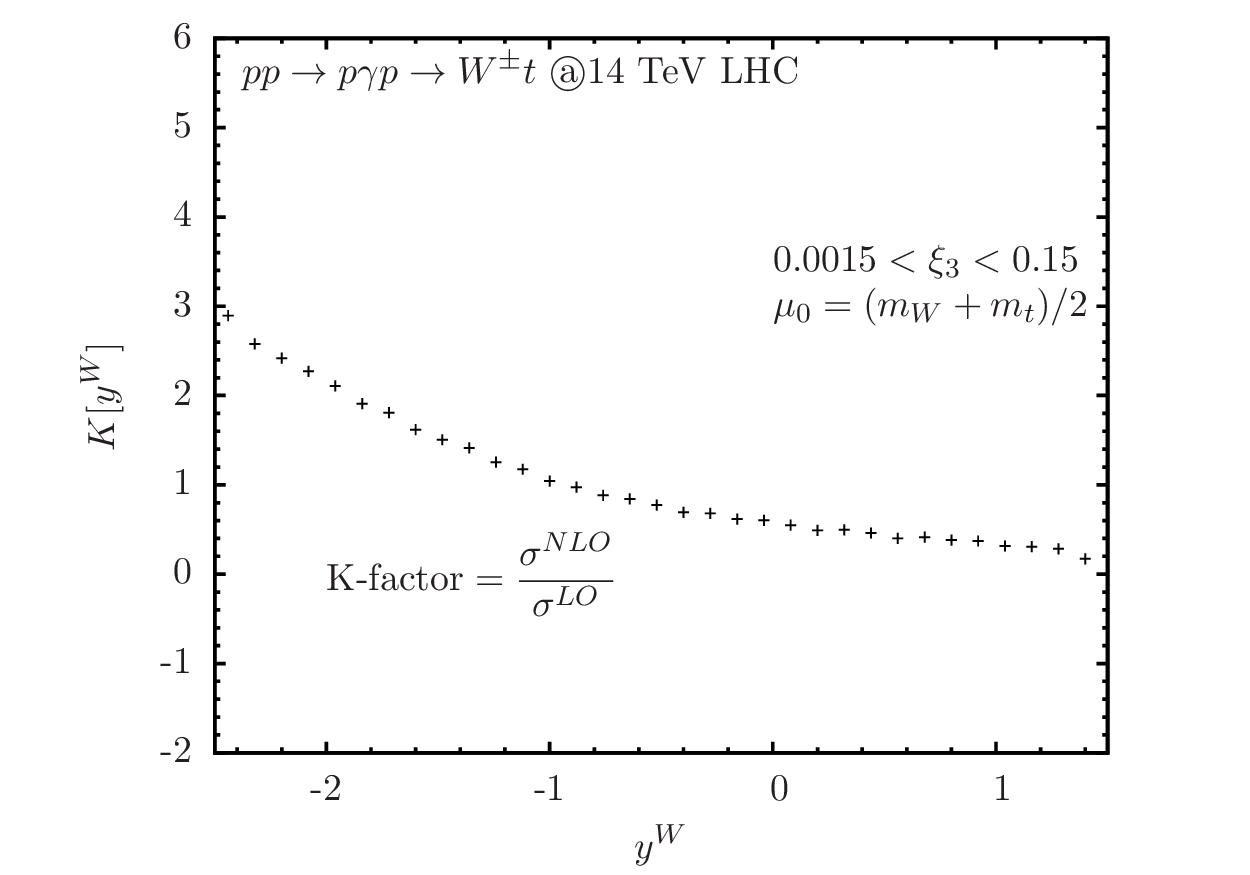}
\par
\end{centering}}

\par
\end{centering}
 \caption{\label{fig10}
The LO (solid curves) and NLO (dotted curves) Rapidity (y) distribution of W [left panel]
and K-factor [right panel] for the process $\rm pp\rightarrow p\gamma p\rightarrow pWt+Y$
at 14 TeV LHC with $\rm \mu=\mu_0=(m_W+m_t)/2$, $\rm \delta_s=10^{-4}$ and $\rm \delta_c=\delta_s/100$.
The experimental detector acceptances are $0.0015<\xi_1<0.5$(a), $0.1<\xi_2<0.5$(b), $0.0015<\xi_3<0.15$(c).}
\end{figure}

\begin{figure}[hbtp]
\begin{centering}
\subfloat[$0.0015<\xi_1<0.5$]
{\begin{centering}
\includegraphics[scale=0.6]{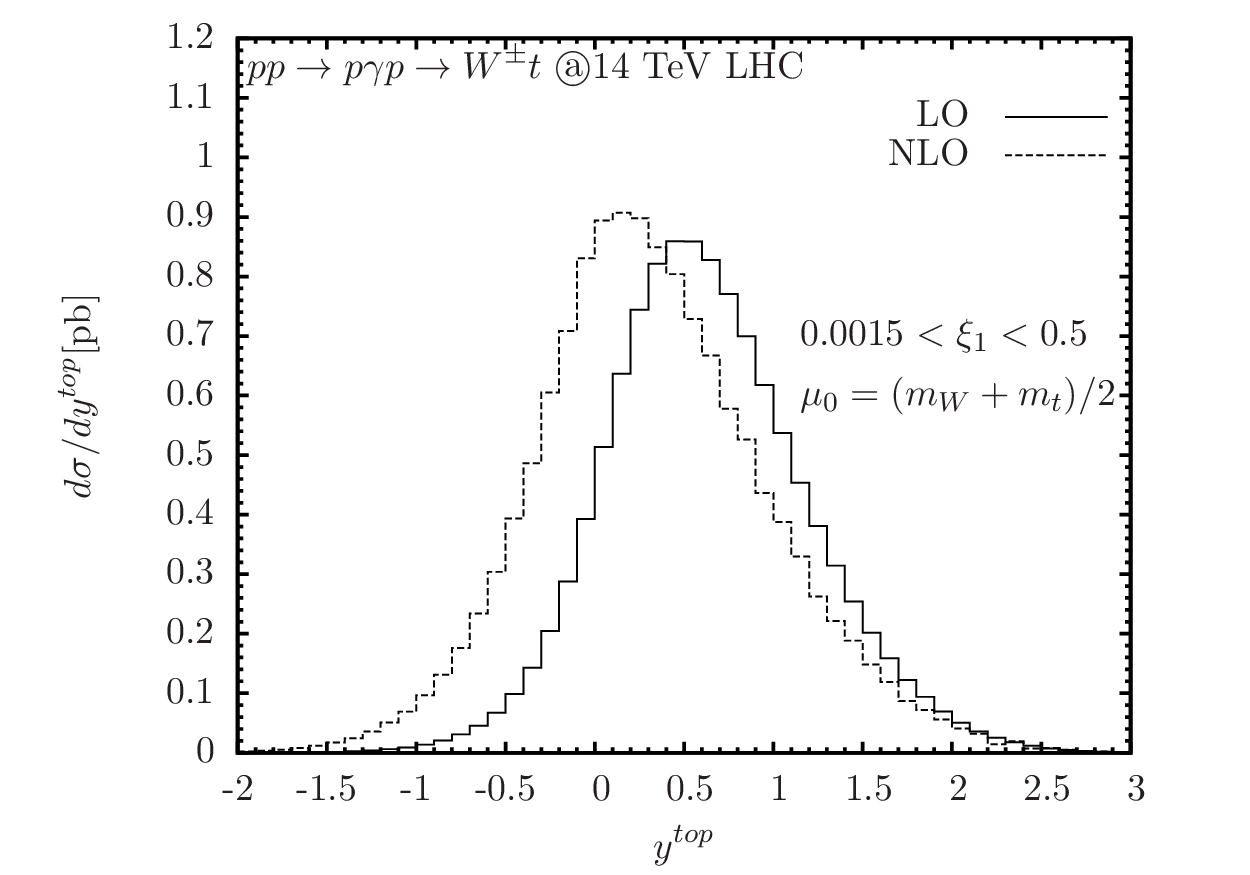}
\includegraphics[scale=0.6]{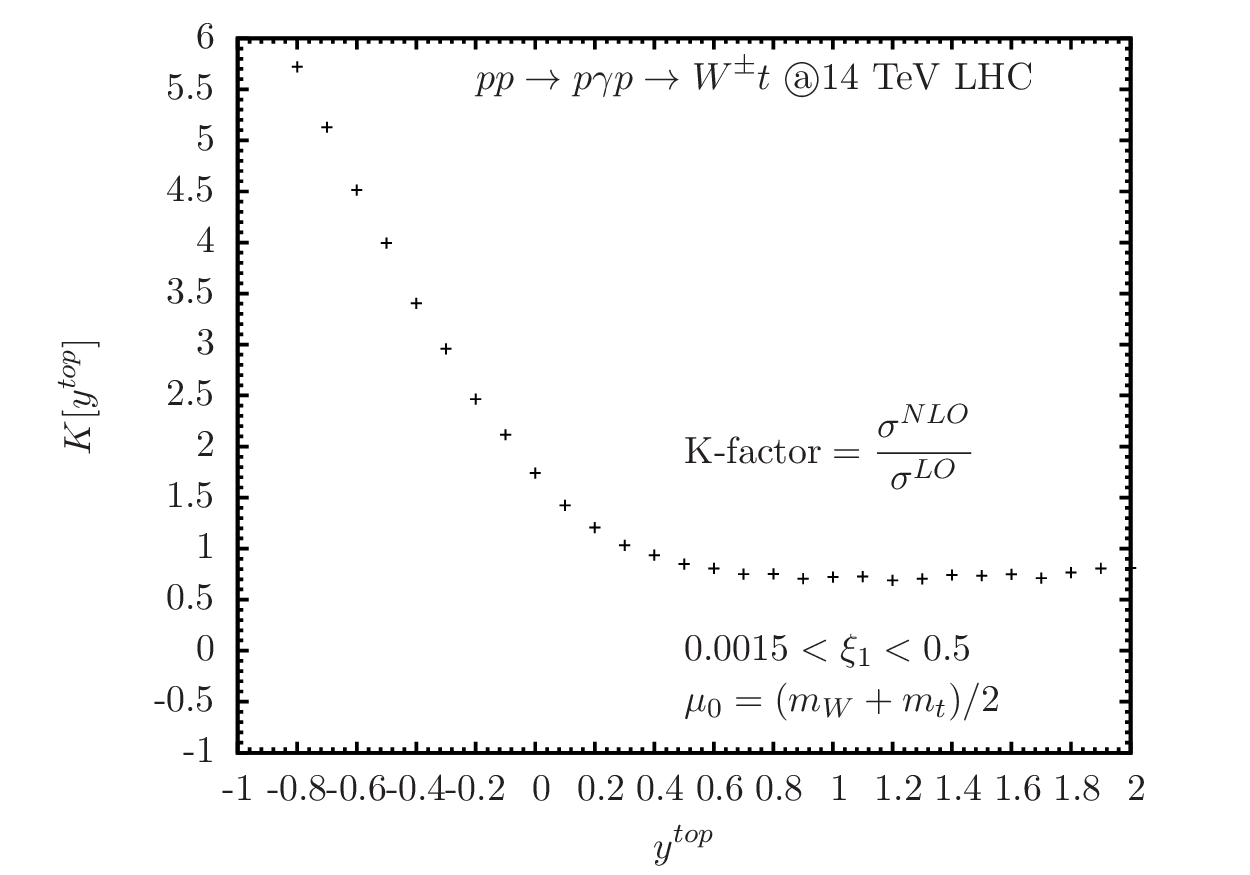}
\par\end{centering}}

\subfloat[$0.1<\xi_2<0.5$]
{\begin{centering}
\includegraphics[scale=0.6]{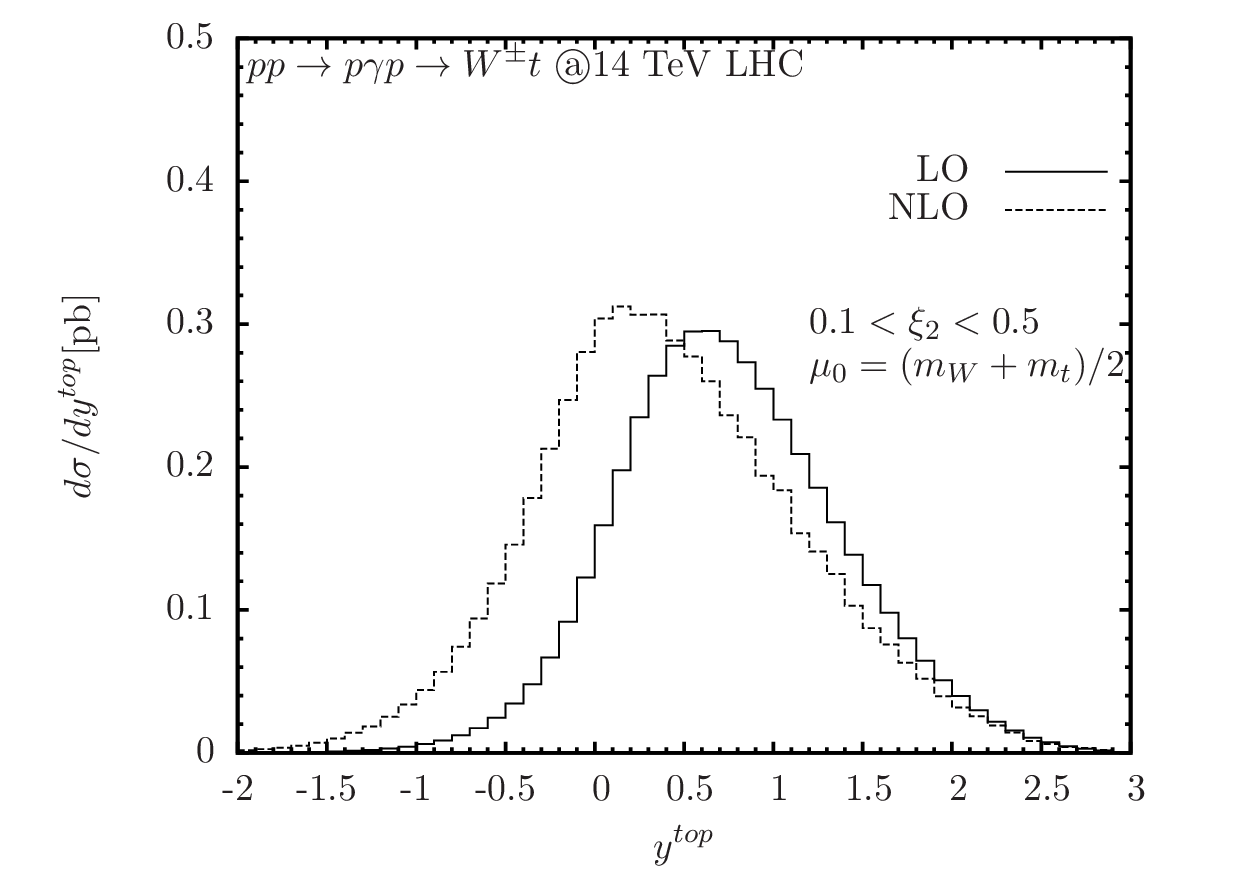}
\includegraphics[scale=0.6]{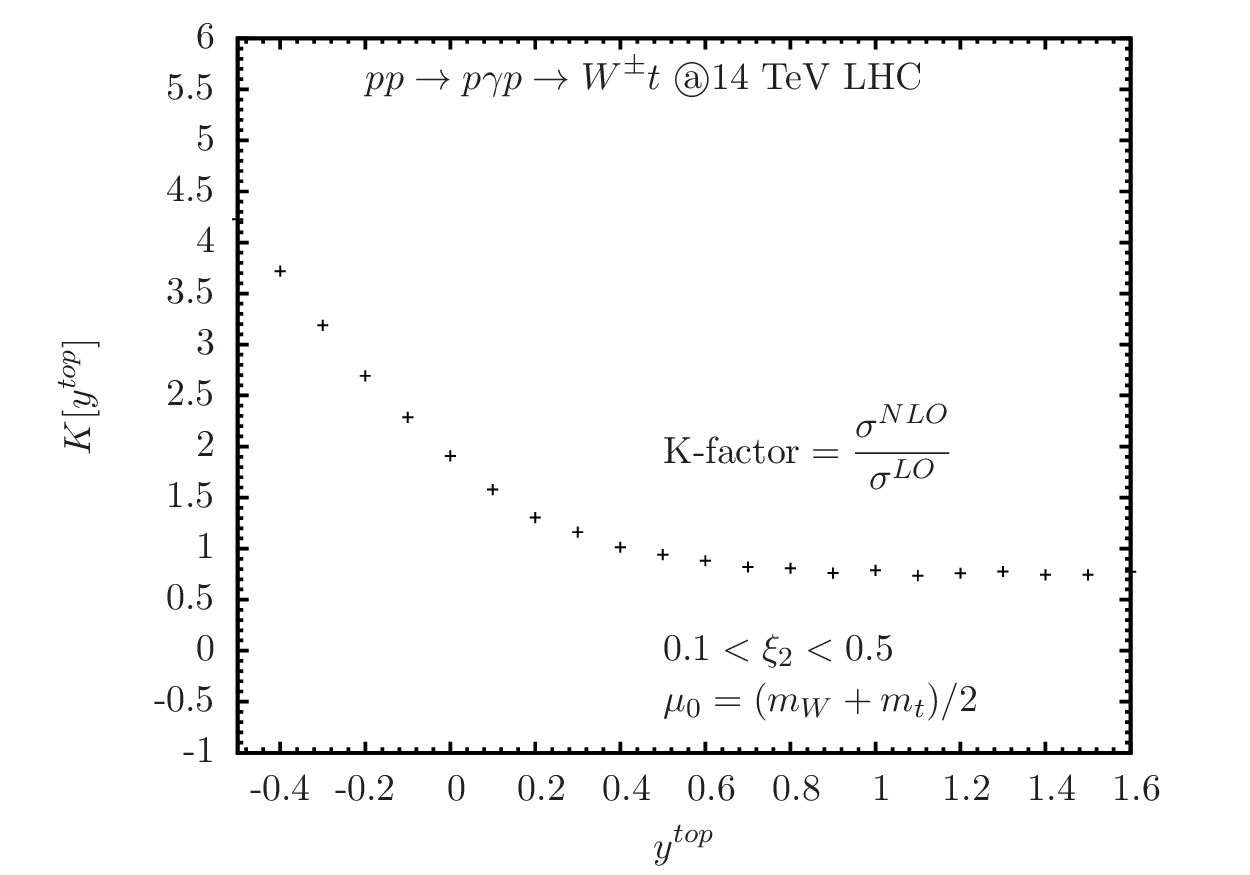}
\par
\end{centering}}

\subfloat[$0.0015<\xi_3<0.15$]
{\begin{centering}
\includegraphics[scale=0.6]{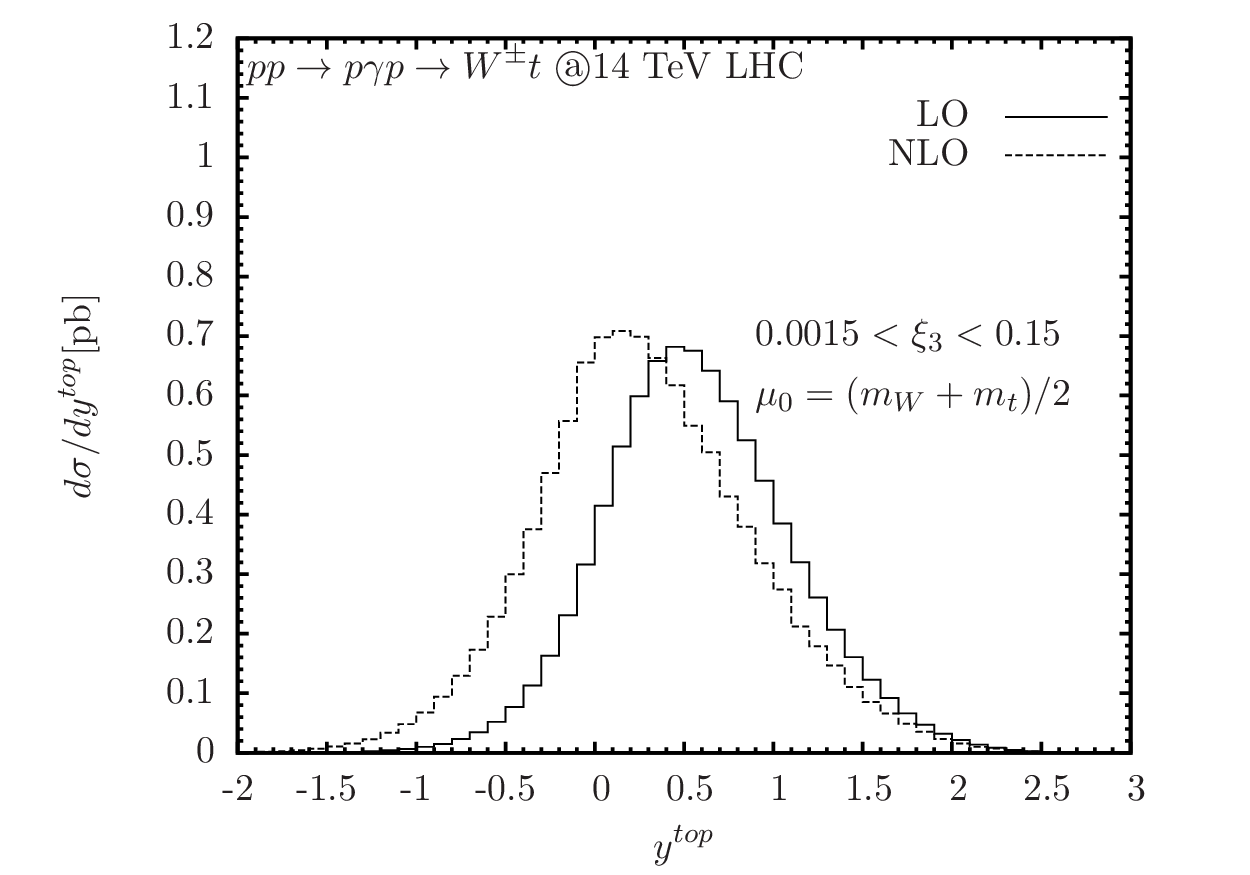}
\includegraphics[scale=0.6]{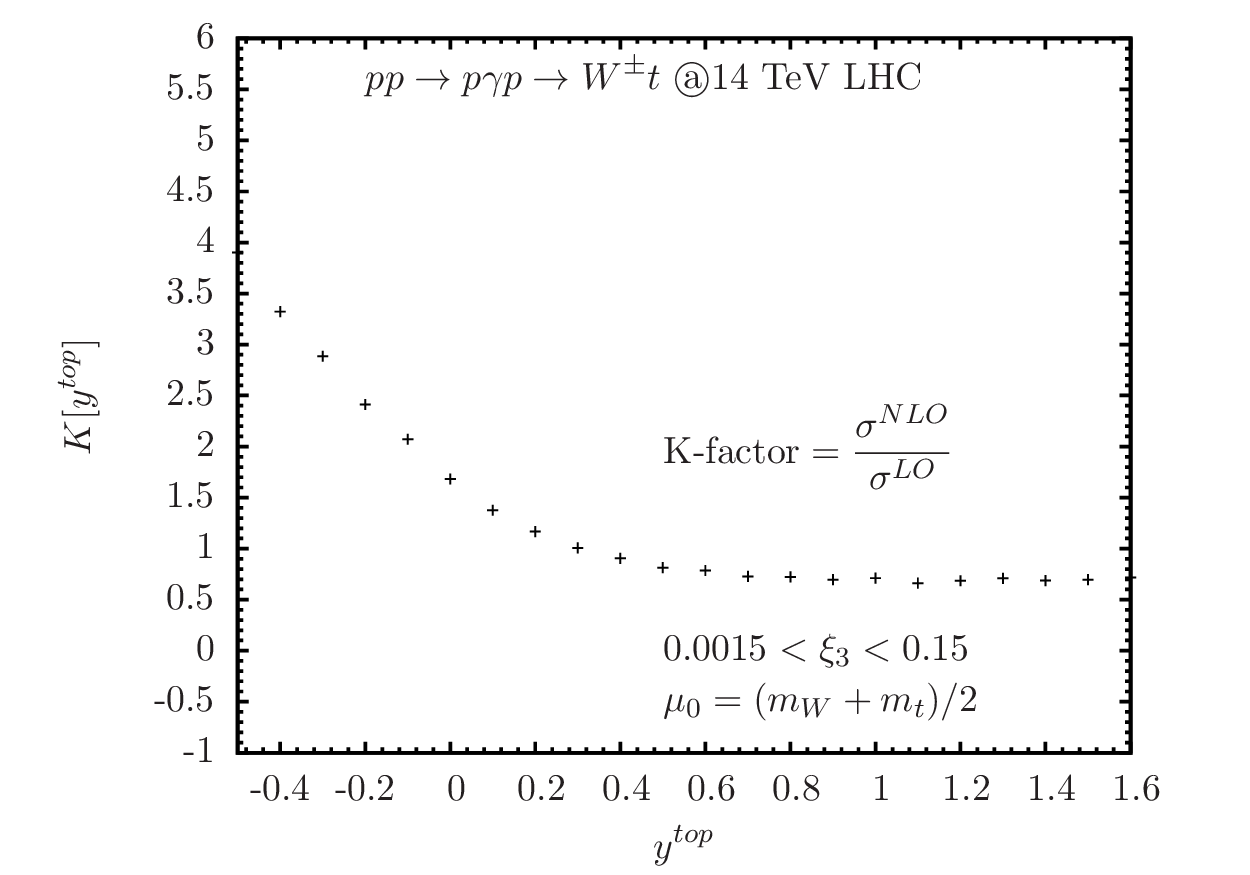}
\par
\end{centering}}

\par
\end{centering}
 \caption{\label{fig11}
The LO (solid curves) and NLO (dotted curves) Rapidity (y) distribution of top [left panel]
and K-factor [right panel] for the process $\rm pp\rightarrow p\gamma p\rightarrow pWt+Y$
at 14 TeV LHC with $\rm \mu=\mu_0=(m_W+m_t)/2$, $\rm \delta_s=10^{-4}$ and $\rm \delta_c=\delta_s/100$.
The experimental detector acceptances are $0.0015<\xi_1<0.5$(a), $0.1<\xi_2<0.5$(b), $0.0015<\xi_3<0.15$(c).}
\end{figure}

Rapidity distributions for the W boson and top quark have been
presented in Fig.\ref{fig10} and Fig.\ref{fig11}. As can be seen the NLO corrections can
shift the LO rapidity but in different ways for both W boson
and top quark. For the W boson the distribution $\rm y^{W}$,
QCD NLO corrections shift the LO peak range into different y and enhance them.
For the top quark rapidity distributions there is not much enhancement can
be found, instead, the rapidity values
where they peaked shifts.
Same behaviors but different values can be found for the other choices of
$\xi_2$ and $\xi_3$ as can be see in Fig.\ref{fig10}(a-c) and Fig.\ref{fig11}(a-c),respectively.
Their corresponding K-factors are present in the right panels.
The reduction can be found for the K-factor when $\rm y^W$ increase from -4 to -1.
With $\rm y^W < -0.8$ NLO correction enhance the LO predictions while
reverse in the range $\rm y^W \geq -0.8$. This behavior is the same for
all three value of forward detector acceptances $\xi_1$, $\xi_2$ and $\xi_3$.
For $\rm y^{top}$, this value is around 0.2. That is to say, for
$\rm y^{top} < 0.2$ NLO correction enhance the LO predictions while
reverse in the range $\rm y^{top} \geq 0.2$.

\begin{figure}[hbtp]
\vspace{-0.2cm}
\centering
\includegraphics[scale=0.8]{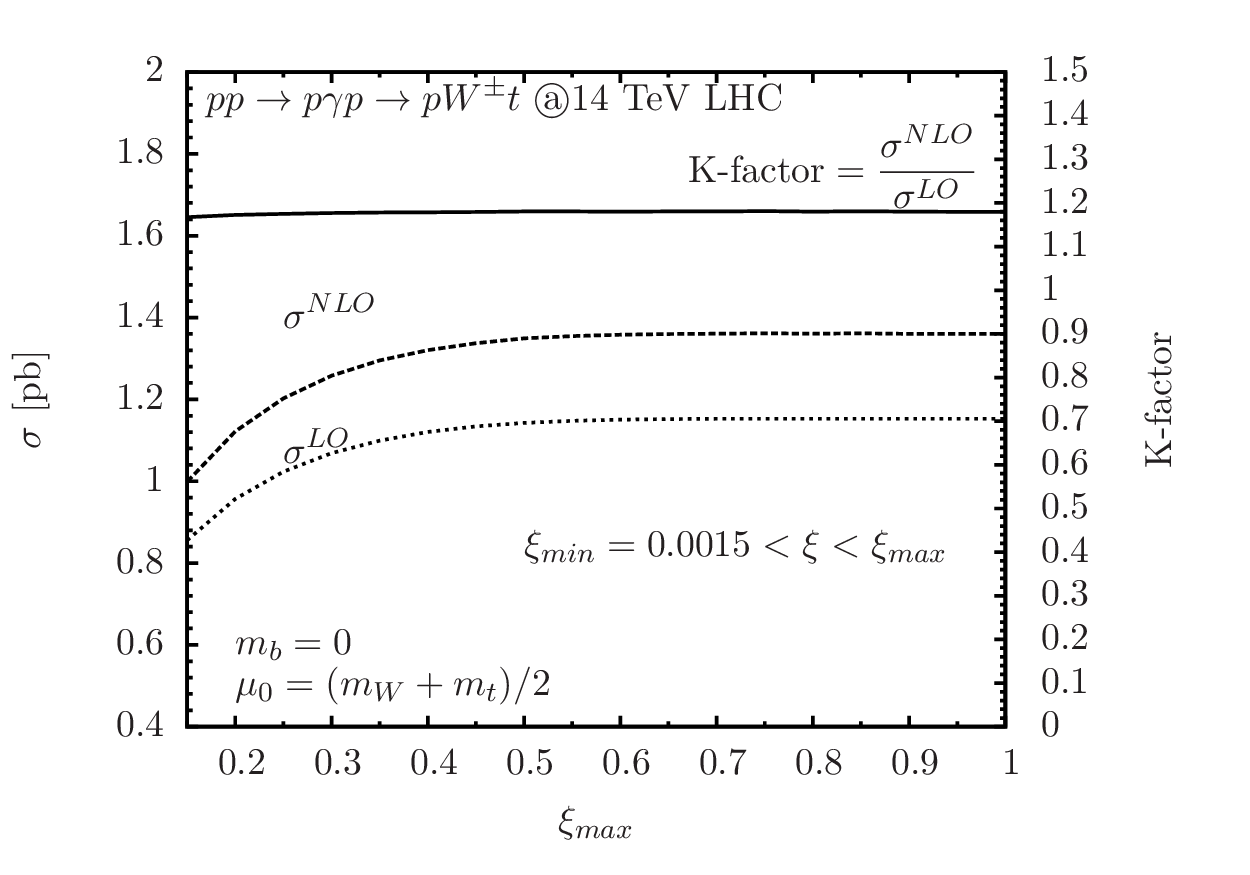}
\caption{\label{fig12}
Cross sections for LO and NLO predictions for $\rm pp\rightarrow p\gamma p\rightarrow pWt+Y$
as well as the the K-factor as functions of different values of $\rm \xi_{max}$ at 14 TeV LHC.
Here we fix $\rm \xi_{min}=0.0015$ and take $\rm \xi_{max}$ as a running parameter from 0.15 to 1.
The dotted, dashed and the solid line correspond to the LO, NLO predictions and K-factor,
respectively. $\rm \mu=\mu_0=(m_W+m_t)/2$, $\rm \delta_s=10^{-4}$ and $\rm \delta_c=\delta_s/100$. }
\end{figure}

In Fig.\ref{fig12} we fix $\rm \xi_{min}=0.0015$ and take $\rm \xi_{max}$ as a running parameter
from 0.15 to 1. The LO and NLO cross sections as well as the the K-factor defined as
$\rm \sigma^{NLO}/\sigma^{LO}$ are presented as functions of different values of $\rm \xi_{max}$.
The dotted, dashed and solid lines correspond to the LO, NLO predictions and K-factor,
respectively. We can find that in the range $\rm 0.0015<\xi_{max}<0.5$, both LO and NLO predictions
rely on the detector acceptances while in the region $\rm \xi_{max}>0.5$, little contributions will
shift the LO and NLO cross sections. No matter how the detector acceptances changes, the ratio of
$\rm \sigma^{NLO}$ to $\rm \sigma^{LO}$ does not change much where a typical value of K-factor equal
1.1808 in the massless assumption, leading the QCD NLO corrections up to
$18.08\%$ related to the LO predictions with our chosen parameters.

\section{Summary}

In this work, we present the precise production of Single Top and W boson
associated photoproduction up to NLO QCD level through the main reaction
$\rm pp\rightarrow p\gamma p \rightarrow pW^\pm t+Y$ at the future 14 TeV
Large Hadron Collider (LHC) for the first time,
assuming a typical LHC multipurpose forward detector.
We use the Five-Flavor-Number Schemes (5FNS) through the whole calculation
while treat the initial state b quark as massless.
This is the most important two body final state
single top production channel at the $\rm \gamma p$ collision.
By detecting this process we can certainly in analyses aiming at top quark
electrical charge, top quark mass prediction, and the CKM matrix element
$\rm |V_{tb}|$ and give complementary information for normal pp collisions.
In this paper, we have employed equivalent photon approximation (EPA) for
the incoming photon beams and performed detailed analysis for various forward
detector acceptances ($\xi$). We analyse their impacts on both the total cross section,
renormalization/factorization scale $\mu$ dependence and some key distributions.
Our results show that: QCD NLO corrections can reduce the
factorization and renormalization scale uncertainty
correspond to their LO predictions.
They can enhance the transverse momentum ($\rm p_T^{W^\pm,top}$)
distributions and shift the LO predictions in different ways for $\rm y^{W^\pm}$ and $\rm y^{top}$,
leading some interesting behaviors and the crucial importance of considering the QCD NLO corrections.
The typical QCD K-factor value in massless b quark scheme are
1.1808 for CMS-TOTEM forward detectors with $0.0015<\xi_1<0.5$,
1.2139 for CMS-TOTEM forward detectors with $0.1<\xi_2<0.5$ and
1.1673 for AFP-ATLAS forward detectors with $0.0015<\xi_3<0.15$, respectively,
with our chosen parameters.

\vskip 10mm
\par
\noindent{\large\bf Acknowledgments:}
Sun Hao thanks Dr. Inanc Sahin for his kindness to provide invaluable comments,
thanks Dr. RenYou Zhang for his strong support and thanks Dr. ChongXing Yue, Lei Guo, ShaoMing Wang for
useful discussions. Project supported by the National Natural Science Foundation of China
(Grant No. 11205070, 11105083, 11035003, 11375104, 11105036),
by Shandong Province Natural Science Foundation (Grant No. ZR2012AQ017),
and by the Fundamental Research Funds for the Central Universities (No. DUT13RC(3)30).

\end{document}